\documentclass[prd,twocolumn,showpacs,showkeys,preprintnumbers,floatfix,nofootinbib,superscriptaddress]{revtex4-1}
\usepackage{amsfonts} 
\usepackage{amssymb} 
\usepackage{amsmath} 
\usepackage{graphicx} 
\usepackage{subfigure} 
\usepackage{array} 
\usepackage{dcolumn} 
\usepackage{bm} 
\usepackage{latexsym} 
\usepackage{longtable} 
\usepackage{hyperref} 
\usepackage{bbold}
\usepackage{color}
\usepackage{multirow}
\usepackage{comment}

\graphicspath{{./figs/}}

\newcommand{\MeV}{\mathop{\rm MeV}\nolimits}

\newcommand{\fm}{\mathop{\rm fm}\nolimits}

\newcommand{\vect}[1]{\ensuremath{\boldsymbol{#1}}}
\newcommand{\vprp}[1]{\vect{#1}_{\mathrm T}}
\newcommand{\vv}{\vect{v}}
\newcommand{\vb}{\vect{b}}

\newcommand{\bvec}{b}

\newcommand{\mN}{m_N}

\newcommand{\zetahat}{{\hat \zeta}}

\begin{document}

%
\title{Nucleon Transverse Momentum-dependent Parton Distributions
in Lattice QCD: Renormalization Patterns and Discretization Effects}
\author{Boram~Yoon}
  \affiliation{Los Alamos National Laboratory, Theoretical Division T-2, Los Alamos, NM 87545, USA}

\author{Michael~Engelhardt}
  \affiliation{Department of Physics, New Mexico State University, Las Cruces, NM 88003-8001, USA}

\author{Rajan~Gupta}  
  \affiliation{Los Alamos National Laboratory, Theoretical Division T-2, Los Alamos, NM 87545, USA}

\author{Tanmoy~Bhattacharya}
  \affiliation{Los Alamos National Laboratory, Theoretical Division T-2, Los Alamos, NM 87545, USA}

\author{Jeremy~R.~Green}
  \affiliation{NIC, Deutsches Elektronen-Synchrotron, 15738 Zeuthen, Germany}



\author{Bernhard~U.~Musch}
  \affiliation{Institut f\"ur Theoretische Physik, Universit\"at
  Regensburg, 93040 Regensburg, Germany}

\author{John~W.~Negele}
  \affiliation{Center for Theoretical Physics, Massachusetts Institute of Technology, Cambridge, Massachusetts 02139, USA}

\author{Andrew~V.~Pochinsky}
  \affiliation{Center for Theoretical Physics, Massachusetts Institute of Technology, Cambridge, Massachusetts 02139, USA}

\author{Andreas~Sch\"afer}
  \affiliation{Institut f\"ur Theoretische Physik, Universit\"at
  Regensburg, 93040 Regensburg, Germany}

\author{Sergey~N.~Syritsyn}
  \affiliation{Department of Physics and Astronomy, Stony Brook University, Stony Brook, NY 11794, USA}
  \affiliation{RIKEN/BNL Research Center, Brookhaven National Laboratory, Upton, NY 11973, USA}

%
%
%
\preprint{LA-UR-17-24472}
%
\pacs{11.15.Ha, 
      12.38.Gc, 
      13.60.Hb  
}
\keywords{Lattice QCD, hadron structure, TMD distribution functions}
\date{\today}
\begin{abstract}
Lattice QCD calculations of transverse momentum-dependent parton
distribution functions (TMDs) in nucleons are presented, based on the
evaluation of nucleon matrix elements of quark bilocal operators with
a staple-shaped gauge connection. Both time-reversal odd effects,
namely, the generalized Sivers and Boer-Mulders transverse momentum
shifts, as well as time-reversal even effects, namely, the generalized
transversity and one of the generalized worm-gear shifts are
studied. Results are obtained on two different $n_f = 2+1$ flavor
ensembles with approximately matching pion masses but very different
discretization schemes: domain-wall fermions (DWF) with lattice
spacing $a=0.084$~fm and pion mass 297~MeV, and Wilson-clover fermions
with $a=0.114$~fm and pion mass 317~MeV. Comparison of the results on
the two ensembles yields insight into the length scales at which
lattice discretization errors are small, and into the extent to which
the renormalization pattern obeyed by the continuum QCD TMD operator
continues to apply in the lattice formulation. For the studied TMD
observables, the results are found to be consistent between the two
ensembles at sufficiently large separation of the quark fields within
the operator, whereas deviations are observed in the local limit and
in the case of a straight link gauge connection, which is relevant to
the studies of parton distribution functions. Furthermore, the lattice
estimates of the generalized Sivers shift obtained here are confronted
with, and are seen to tend towards, a phenomenological estimate
extracted from experimental data.
\end{abstract}
\maketitle
%
%
%
%
\section{Introduction}
\label{sec:into}

An important aspect of nucleon internal dynamics is the
three-dimensional momentum carried by quarks, comprising not only the
longitudinal momentum fraction $x$ encoded in standard parton
distribution functions, but also momentum in the transverse plane. It
is characterized by transverse momentum-dependent parton distribution
functions (TMDs). TMDs enter in, for example, angular asymmetries
measured in semi-inclusive deep inelastic scattering (SIDIS) processes
of electrons off nucleons. Depending on the polarizations of the
nucleon and the struck quark, a number of correlations can be studied,
including the time-reversal odd (T-odd) effects encoded in the Sivers
and Boer-Mulders TMDs. These are a consequence of final state
interactions in the SIDIS process, and analogously manifest themselves
via initial state interactions in the Drell-Yan (DY) process. TMDs are
a focus of experiments at the JLab 12 GeV facility and at RHIC, and
constitute an important component of the motivation for the proposed
electron-ion collider (EIC).

%
%

To obtain first-principles, nonperturbative input for the theoretical
study of TMDs, a method to evaluate TMD observables in Lattice QCD has
been developed and explored in
\cite{Hagler:2009mb,Musch:2010ka,Musch:2011er}.  In the present work,
we report results obtained on two gauge
ensembles at approximately matching pion masses, but with substantially
differing fermion discretization schemes: One is a 2+1-flavor RBC/UKQCD
domain wall fermion ensemble with lattice spacing $a=0.084\fm$ and pion
mass $297\MeV$ \cite{Allton:2008pn}, the other is a 2+1-flavor isotropic
clover fermion ensemble generated by R.~Edwards, B.~Jo\'o and
K.~Orginos~\cite{JLAB:2016} with lattice spacing $a=0.114\fm$ and pion
mass $317\MeV$. Aside from being located closer to the physical pion
mass than the aforementioned previous investigations, the availability
of data on these two separate ensembles allows us to investigate 
two specific facets of the lattice TMD calculational scheme pursued
here; namely, discretization effects and the renormalization of the
quark bilocal operators used in the definition and evaluation of TMDs,
laid out in detail in section \ref{sec:setup}.

The composite operator used to extract TMDs consists of a quark and an
antiquark field connected by an intrinsically nonlocal gauge
connection -- a path ordered product of gauge links. The divergences
associated with the quantum fluctuations of the latter are absorbed
into a multiplicative ``soft factor'' in the continuum QCD scheme
developed in
Ref.~\cite{Aybat:2011ge,Collins:2011zzd}.\footnote{Throughout this
  paper, the label ``soft factor'' denotes both soft and collinear
  divergences.}  In addition, renormalization factors are attached to
the quark fields. This multiplicative nature of renormalization in
continuum QCD is central to the construction of the TMD observables
considered here, in which the renormalization factors are canceled by
forming suitable ratios. Whether this multiplicative nature of
renormalization carries over into the lattice framework is, however, a
point which demands further investigation. One possible manifestation
of the violation of multiplicativity would be if results for the
aforementioned TMD ratios vary with the lattice discretization scheme,
and the difference persists as the lattice spacing is taken to zero.

The availability of lattice TMD data on two ensembles with
approximately matching pion masses, but differing discretizations
provides an opportunity for an empirical test of the universality
of TMD ratios. This is a primary focus of the present work. One
would expect that the lattice operators approximate the continuum
operators well at finite physical extent, and that results obtained
on the two ensembles therefore match. On the other hand,
the local limit may exhibit additional ultraviolet divergences, as
well as signatures of operator mixing attributable to the breaking
of continuum symmetries, such as rotational symmetry and, in the
case of clover fermions, chiral symmetry. It is well known that in
the local limit, renormalization constants of composite operators
become dependent on the Dirac structure, whereas the soft
factors and quark wave function renormalizations used to renormalize
the nonlocal TMD operators do not depend on the Dirac structure. The working 
assumption underlying the construction of the TMD observables considered
in this work is that, at large enough separation, the renormalization
factors become independent of the Dirac structure and cancel in ratios.
Comparing the results obtained on the two ensembles is expected to
uncover whether, at what length scales, and under what conditions
this assumption holds, and whether any signatures of deviations
from this simple renormalization pattern can be detected.

This paper is organized as follows: Sec.~\ref{sec:setup} lays out
the definition of TMDs and the construction of TMD ratio observables
in which multiplicative soft factors and renormalization constants
cancel. Particulars of the Lattice QCD evaluation of these quantities
are given in Sec.~\ref{sec:latt}. Results for the Sivers shift,
Boer-Mulders shift, transversity $h_1$ and the $g_{1T} $ worm-gear shift 
are given in Sec.~\ref{sec:res}. Some results pertinent to the calculation of 
parton distribution functions (PDFs) are given in Sec.~\ref{sec:straight}. A comparison of our estimate of the
generalized Sivers shift with that extracted from SIDIS experiments is
presented in Sec.~\ref{sec:comp-sivers}. Conclusions are given
in Sec.~\ref{sec:sum}.

\section{Construction of TMD observables}
\label{sec:setup}

%
\begin{figure}[tb]
  \includegraphics[width=0.96\linewidth]{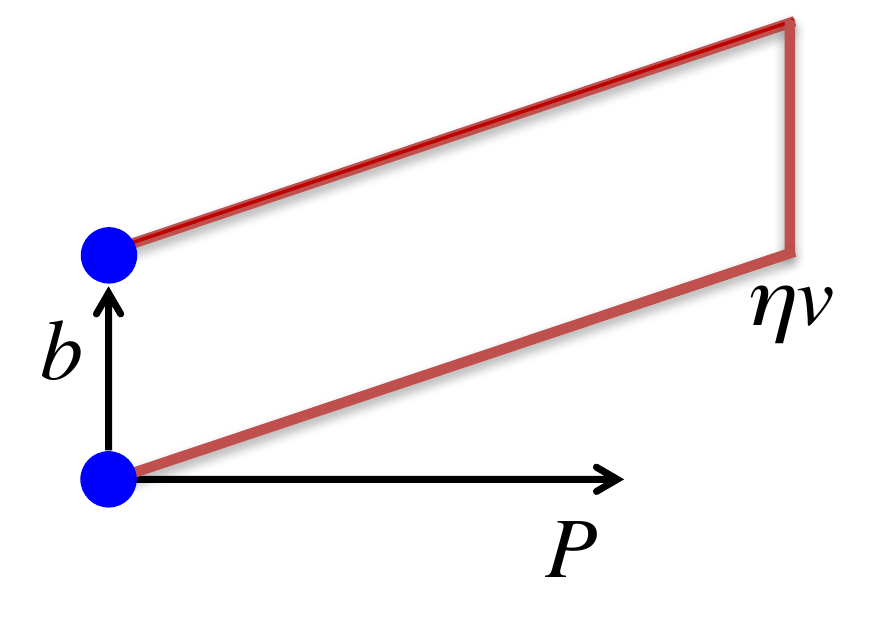}
\caption{Illustration of the TMD operator with staple-shaped gauge
  connection. The four-vectors $v$ and $P$ give the direction of the
  staple and the momentum, while $b$ defines the separation between
  the quark operators. The values of these variables used in the
  lattice calculation are given in
  Table~\protect\ref{tab:meas_params}.  In the present calculation,
  $b\cdot P = b\cdot v =0$ is chosen, corresponding to evaluating the
  first moment of the TMDs with respect to the quark momentum fraction $x$.}
\label{fig:TMDop}
\end{figure}

The calculational scheme employed to arrive at lattice TMD observables
has been laid out in detail in~\cite{Musch:2011er}, cf.~also
\cite{Hagler:2009mb,Musch:2010ka}. The
following synopsis emphasizes, in particular, how multiplicative soft
factors enter the scheme, and the consequent construction of TMD ratios
in which these factors cancel. TMDs are derived from the fundamental
correlator
\begin{eqnarray}
& \widetilde{\Phi }^{[\Gamma ]}_{\mbox{\scriptsize unsubtr.} }
(b,P,S,\ldots ) \hspace{5cm} & \label{spacecorr} \\
& \hspace{1.2cm} \equiv \frac{1}{2} \langle P,S | \ \bar{q} (0) \
\Gamma \ {\cal U} [0,\eta v, \eta v+b,b] \ q(b) \ |P,S\rangle &
\nonumber
\end{eqnarray}
where the subscript ``unsubtr.'' indicates that no provision has been
made yet to absorb ultraviolet and soft divergences into appropriate
renormalization factors. The nucleon states are characterized by the
longitudinal momentum $P$ and the spin $S$. We will consistently use
the tilde, as in $\widetilde{\Phi }$, to denote position space
correlation functions and the same symbols without the tilde for their
Fourier transforms. The quark fields, separated by a displacement $b$,
are connected by the gauge connection ${\cal U} [0,\eta v, \eta v+b,b]$,
the arguments of which denote space-time positions connected by path-ordered
products of gauge links approximating straight Wilson lines. The full
gauge connection thus has the shape of a staple, with the direction of
the staple encoded in the vector $v$, and its length in the scalar
$\eta $ as shown in Fig.~\ref{fig:TMDop}. We are specifically interested
in the limit $|\eta | \to \infty$, in which this gauge connection represents
gluon exchange in semi-inclusive deep inelastic scattering (SIDIS) and
Drell-Yan (DY) processes. The directions of the staples in the two cases
are opposite to one another; in the SIDIS case, the staple-shaped gauge
connection incorporates final state interactions of the struck quark,
whereas in the DY case, it incorporates initial state interactions.

An additional important specification regarding the concrete
choice of the staple direction $v$ is needed. In the definition
of TMDs, $v$ is taken to have no transverse component, $\vprp{v} =0$.
Furthermore, in a hard scattering process, the rapidity difference
between the incoming hadron and the struck quark is very large, and a
natural choice for $v$ would therefore be a light-cone vector.
However, such a choice leads to severe rapidity divergences
beyond tree level \cite{Collins:1981uw}, which are regulated by taking
$v$ off the light cone into the space-like region. Consequently,
TMDs depend on an additional Collins-Soper type parameter $\zetahat $
characterizing how close $v$ is to the light cone,
\begin{equation}
\zetahat = \frac{v\cdot P}{\sqrt{\vert v^2 \vert}\sqrt{P^2} } \ .
\label{zetahatdef}
\end{equation}
The light-cone limit corresponds to $\zetahat \rightarrow \infty $.
Note that this limit can be approached even with a purely spatial choice
of $v$, as used in lattice calculations, if the spatial momentum
$\vect{P} $ is chosen large. In practice, Lattice QCD calculations only
access a fairly limited range of $\zetahat $ because of limitations in 
performing simulations at large $\vect{P} $; ultimately, one aims to connect to the
region of sufficiently large values in which perturbative evolution
equations become applicable \cite{Collins:1981uk, Aybat:2011ge}.
A dedicated lattice study of the $\zetahat $-scaling of a TMD
observable was performed in \cite{Engelhardt:2015xja}. A perspective
for extending lattice TMD calculations to higher nucleon momenta, and
therefore higher $\zetahat $, is given by the recently developed
momentum smearing method described in \cite{Bali:2016lva}.

To regulate the TMD correlator defined in Eq.~\eqref{spacecorr}, one
considers subtracted correlation functions
\cite{Collins:2011zzd,Aybat:2011ge}
\begin{eqnarray}
&\widetilde{\Phi }^{[\Gamma ]}_{\mbox{\scriptsize subtr.} }
(b,P,S,\ldots )  \hspace{5.5cm} &  \label{eq:subtracted} \\ 
& \hspace{1.2cm} =  \widetilde{\Phi }^{[\Gamma ]}_{\mbox{\scriptsize unsubtr.} }
(b,P,S,\ldots ) \cdot {\cal S} \cdot Z_{\rm TMD} \cdot Z_2  \,, & \nonumber
\end{eqnarray}
in which divergences have been absorbed into three separate factors:
${\cal S}$ regulates the soft and collinear divergences associated
with the gauge connection, $Z_2$ is the quark field renormalization
factor, and the rest, $Z_{\rm TMD}$, contains the dependence on the
specific tensor structure of the TMD operator under consideration. As
discussed in Refs.~\cite{Collins:2011zzd,Aybat:2011ge}, the factor
${\cal S}$ is defined only in terms of Wilson lines, and $Z_2$ is also
independent of the particular choice of the TMD operator. In
Eqs.~\eqref{eq:GSS},~\eqref{eq:BMS},~\eqref{eq:transversity},
and~\eqref{eq:WGS}, we define the four observables we calculate as
ratios, in which the unpolarized TMD moment $\tilde f_1^{[1](0)}$,
cf.~Eqs.~\eqref{tmd1} and~\eqref{eq:fdef}, is used to define the
denominator. The reason for studying ratios rests on the assumption
that the full renormalization $\cal Z$ continues to factor in the
lattice formulation, i.e., the renormalization pattern given in
Eq.~\eqref{eq:subtracted} with $ {\cal Z} = {\cal S} \cdot Z_{\rm TMD}
\cdot Z_2$ also holds on the lattice. In that case, the two factors
${\cal S} \cdot Z_2 $ would cancel in the ratios. The additional
assumption is that for finite physical separation, $b$, the factor
$Z_{\rm TMD} $ also becomes independent of the spin ($\gamma $-matrix)
structure of the TMD operator, and therefore it also cancels in the
ratio. Note that, at finite lattice cutoff, there is no hard
separation between the local limit and finite physical distances, and
a smooth transition in behavior occurs over several lattice
spacings. A similar multiplicative renormalization is used to regulate
the operator used in studies of
PDFs~\cite{Ishikawa:2016znu,Chen:2016fxx,Chen:2017mzz,Alexandrou:2017huk}. 

The analysis of TMDs in the continuum is in terms of the subtracted
correlation function
$\widetilde{\Phi }^{[\Gamma ]}_{\mbox{\scriptsize subtr.} }$, which upon
Fourier transformation yields the momentum space correlator
\begin{eqnarray}
& \Phi^{[\Gamma ]} (x,\vect{k}_{T},P,S,\ldots ) \hspace{5cm}
& \label{momcorr} \\
& \hspace{0.7cm}
=\int \frac{d^2 \vprp{b} }{(2\pi )^2 } \int \frac{d(b\cdot P)}{2\pi P^{+} }
e^{ix(b\cdot P)-i\vprp{b} \cdot \vprp{k} } \left.
\widetilde{\Phi }^{[\Gamma ]}_{\mbox{\scriptsize subtr.} } 
\right|_{b^{+} =0} \nonumber
\end{eqnarray}
in which the suppressed momentum component $k^{-} $ is integrated over,
leading to the specification $b^{+} =0$. The transverse components $\vprp{b} $
of the quark separation $b$ are Fourier conjugate to the quark transverse
momentum $\vprp{k} $, whereas the longitudinal component $b\cdot P$ is
Fourier conjugate to the longitudinal momentum fraction $x=k^{+}/P^{+} $. 
In the present work we restrict to the case $b\cdot P=0$, thus obtaining 
only the integral with respect to $x$ of the correlator $\Phi^{[\Gamma ]} $
and all TMDs derived from it. It is, however, important to note that
lattice calculations can be extended to scan the
$b\cdot P$-dependence\footnote{In a practical calculation,
the range of accessible $b\cdot P$ is limited by the available $b$ and
$P$, $|b\cdot P| \leq |\vect{P} | \sqrt{-b^2 } $, 
leading to an increasing systematic
uncertainty at small $x$.} and obtain, after Fourier transformation,
the $x$-dependence of $\Phi^{[\Gamma ]} $ and the TMDs under consideration. 
Studies of the $b\cdot P$-dependence in the case of straight gauge links
($\eta v=0$) were carried out in Ref.~\cite{Hagler:2009mb,Musch:2010ka}
and a related project to obtain the $x$-dependence of parton distribution
functions (PDFs) has been developed in
Refs.~\cite{Ji:2013dva,Lin:2014zya,Alexandrou:2015rja,Ishikawa:2016znu,Chen:2016fxx,Chen:2016utp,Chen:2017mzz,Alexandrou:2016jqi,Alexandrou:2017huk}. 

At leading twist, Eq.~\eqref{momcorr}
defines eight TMDs as coefficient functions with the parametrization
\begin{eqnarray}
\Phi^{[\gamma^{+} ]} &=& f_1 -
\frac{\epsilon_{ij} \vect{k}_{i} \vect{S}_{j} }{\mN }
f_{1T}^{\perp } \label{tmd1} \\
\Phi^{[\gamma^{+} \gamma^{5} ]} &=& \Lambda g_1
+\frac{\vprp{k} \cdot \vprp{S} }{\mN } g_{1T} \\
\Phi^{[i\sigma^{i+} \gamma^{5} ]} &=& \vect{S}_{i} h_1
+\frac{(2\vect{k}_{i} \vect{k}_{j} -\vprp{k}^2 \delta_{ij} )
\vect{S}_{j} }{2\mN^2 } h_{1T}^{\perp }
\label{tmd3} \\
& & \hspace{0.8cm}
+\frac{\Lambda \vect{k}_{i} }{\mN } h_{1L}^{\perp }
+\frac{\epsilon_{ij} \vect{k}_{j} }{\mN } h_{1}^{\perp }
\nonumber
\end{eqnarray}
where $\mN$ denotes the mass,  $\Lambda $ the helicity and $\vect{S}_T$ the 
transverse spin of the
nucleon. On the lattice, we instead calculate directly the position
space correlation function Eq.~\eqref{spacecorr} using unrenormalized
operators, which, analogous to Eqs.~\eqref{tmd1}--\eqref{tmd3}, can
also be parametrized in terms of Lorentz invariant
amplitudes. Specializing to $b^{+} =0$ and $\vprp{v}=\vprp{P} =0$, and
working again at leading twist, one has
\begin{eqnarray}
\frac{1}{2P^{+} }
\widetilde{\Phi }^{[\gamma^{+} ]}_{\mbox{\scriptsize unsubtr.} }
&=& \widetilde{A}_{2B} +i\mN \epsilon_{ij} \vb_{i} \vect{S}_{j}
\widetilde{A}_{12B} \label{atilde1} \\
\frac{1}{2P^{+} }
\widetilde{\Phi }^{[\gamma^{+} \gamma^{5} ]}_{\mbox{\scriptsize unsubtr.} }
&=& -\Lambda \widetilde{A}_{6B} \\
&+& i [(b\cdot P)\Lambda 
 -\mN (\vprp{b} \cdot \vprp{S} )]
\widetilde{A}_{7B} \nonumber \\
\frac{1}{2P^{+} }
\widetilde{\Phi }^{[i\sigma^{i+} \gamma^{5} ]}_{\mbox{\scriptsize unsubtr.} }
&=& i\mN \epsilon_{ij} \vb_{j} \widetilde{A}_{4B} -\vect{S}_{i}
\widetilde{A}_{9B} \label{atilde3} \\
& & -i\mN \Lambda \vb_{i} \widetilde{A}_{10B} \nonumber \\
& & +\mN [(b\cdot P)\Lambda
-\mN (\vprp{b} \cdot \vprp{S} )] \vb_{i} \widetilde{A}_{11B}
\nonumber
\end{eqnarray}
Note that the Lorentz invariant amplitude combinations
$\widetilde{A}_{iB} $ are suitable linear combinations of the
amplitudes one finds in the most general decomposition, in the absence
of the aforementioned constraints on $b$, $v$ and $P$.  The detailed
decomposition into the various amplitudes as calculated by us on the
lattice is given in Eqs.(16)--(20) of Ref.~\cite{Musch:2011er}.

Clearly, there are parallels between
Eqs.~\eqref{tmd1}-\eqref{tmd3} and
Eqs.~\eqref{atilde1}-\eqref{atilde3}. Since the left-hand sides are
essentially Fourier transforms of one another,
the amplitude combinations $\widetilde{A}_{iB} $
are related to Fourier-transformed TMDs through the following
relations, as explained in detail in \cite{Musch:2011er}:
\begin{eqnarray}
\tilde{f}_{1}^{[1](0)} &=& 2\widetilde{A}_{2B} / {\cal Z}^f \label{rela1} \\
\tilde{g}_{1}^{[1](0)} &=& -2\widetilde{A}_{6B} / {\cal Z}^g \label{rela2} \\
\tilde{g}_{1T}^{[1](1)} &=& -2\widetilde{A}_{7B} / {\cal Z}^g \label{rela3} \\
\tilde{h}_{1}^{[1](0)} &=& -2\left( \widetilde{A}_{9B} -
(\mN^2 b^2 /2) \widetilde{A}_{11B} \right) / {\cal Z}^h \label{rela4} \\
\tilde{h}_{1L}^{\perp [1](1)} &=& -2\widetilde{A}_{10B} / {\cal Z}^h
\label{rela5} \\
\tilde{h}_{1T}^{\perp [1](2)} &=& 4\widetilde{A}_{11B} / {\cal Z}^h 
\label{rela6} \\
\tilde{f}_{1T}^{\perp [1](1)} &=& -2\widetilde{A}_{12B} / {\cal Z}^f  \label{rela7}\\
\tilde{h}_{1}^{\perp [1](1)} &=& 2\widetilde{A}_{4B} / {\cal Z}^h \label{rela8}
\end{eqnarray}
where the superscript ``$[1]$'' indicates that the first Mellin moment
with respect to quark momentum fraction $x$ of the Fourier transform
of a generic TMD $f(x,\vprp{k}^{2} ,\ldots )$ has been
taken~\cite{Engelhardt:2015xja},
\begin{eqnarray}
\tilde{f}^{[m](n)} ( \vprp{\bvec}^2 ,\ldots ) & \equiv &
n!\left( -\frac{2}{\mN^2}\partial_{\vprp{\bvec}^2} \right)^n \
\int_{-1}^{1} dx\, x^{m-1} \cdot
\label{eq:fdef} \\
& & \ \ \ \ \ \ \ \ \ \ \
\cdot \int d^2 \vprp{k} \,
e^{i\vprp{b} \cdot \vprp{k} } f(x,\vprp{k}^{2} ,\ldots ) \ .
\nonumber
\end{eqnarray}

We have introduced three renormalization factors ${\cal Z}^{f,g,h}$ in
Eqs.~\eqref{rela1}-\eqref{rela8} reflecting the tensor structure of
the three TMD operators considered. It is important to note that there
is a different ${\cal Z}^{f,g,h}$ for each staple geometry (soft
factor) and separation $b$. In the generalized Sivers shift, defined
in Eq.~\eqref{eq:GSS}, the two terms, $\tilde
f_{1T}^{\perp[1](1)}(\vprp{\bvec}^2;\ldots)$ and $ {\tilde
  f_1^{[1](0)}(\vprp{\bvec}^2;\ldots)} \,,$ are obtained via
Eqs.~\eqref{rela1} and~\eqref{rela7}, in conjunction with
Eq.~\eqref{atilde1}, from the matrix element of the same operator with
the same value of $b$ and $\eta$. The two terms $\widetilde{A}_{2B}$
and $\widetilde{A}_{12B}$ are isolated, cf.~Eq.~\eqref{atilde1}, by
using different values of $\epsilon_{ij} \vb_{i} \vect{S}_{j}$. Thus
we expect the factor ${\cal Z}^{f}$ to be the same and to cancel in
the ratio. For the Boer-Mulders, transversity and worm-gear shifts
defined in Eqs.~\eqref{eq:BMS},~\eqref{eq:transversity},
and~\eqref{eq:WGS}, the renormalization factor, {\it a priori}, does
not cancel in the ratio even if renormalization is multiplicative. Ignoring 
discretization errors, we would then attribute the difference in results between different fermion
formulations to $Z_{\rm TMD}$. It is at this point that the
assumption underlying lattice calculations stated previously, that at
sufficiently large $b$ in physical units all the $Z_{\rm TMD}$ become independent of the
spin structure of the operator, is important. In that case, the full
renormalization factor would again cancel in the ratios constructed
for fixed but large $b$ and fixed $\eta$, i.e., for the same staple
geometry.

Clearly, such full cancellation will not extend to the $b\rightarrow
0$ limit in general. As a case in point, consider the ratio of matrix
elements of the (isovector) local axial vector and vector currents
within any state $|P,S\rangle$,
\begin{equation}
\left.
\frac{\langle P,S| \bar{q} \gamma^{+} \gamma^{5} q|P,S\rangle }{\langle P,S|
\bar{q} \gamma^{+} q|P,S\rangle } \right|_{ren.} =
\frac{Z_A }{Z_V } \left.
\frac{\langle P,S| \bar{q} \gamma^{+} \gamma^{5} q|P,S\rangle }{\langle P,S|
\bar{q} \gamma^{+} q|P,S\rangle } \right|_{bare}
\label{locrat}
\end{equation}
This ratio is related to, though not identical to, the $b\rightarrow
0$ limit of the ratio in Eq.~\eqref{eq:WGS}; while the denominator of
Eq.~\eqref{eq:WGS} indeed reduces to the vector current for
$b\rightarrow 0$, the numerator corresponds to a higher $\vprp{k}
$-moment of the nonlocal axial vector operator. Nonetheless, if
discrepancies between different fermion discretizations arise for
Eq.~\eqref{locrat}, then they must also be countenanced for the
$b\rightarrow 0$ limit of Eq.~\eqref{eq:WGS}.  Indeed, chiral symmetry
implies $Z_A =Z_V $, such that the renormalization factors in
Eq.~\eqref{locrat} cancel in Lattice QCD as long as one uses a (to a
good approximation) chirally symmetric discretization; two examples
are domain wall and overlap fermions. However, they do not cancel for
clover fermions; for the clover ensemble considered here,
$Z_A/Z_V=1.096(22)$ \cite{Yoon:2016jzj}. A discrepancy between the
unrenormalized ratios obtained in the two fermion discretization
schemes thus arises, over and above that expected from finite lattice
spacing effects alone. Evidence of such a difference in the worm-gear
ratio defined in Eq.~\eqref{eq:WGS} at small $\vprp{b} $ is discussed
in Sec.~\ref{sec:WGS}. In the cases of the generalized Boer-Mulders
shift and the tensor charge, chiral symmetry arguments do not
constrain the ratio $Z_T/Z_V$ of a local tensor to vector operator,
and this ratio can be significantly different for DWF versus clover
fermions at finite lattice spacing $a$. Thus the ratio ${\cal
  Z}^h/{\cal Z}^f$ need not cancel and the results can be
$b$-dependent at small $b$.

Expanding upon a point mentioned above, the $b=0$ limit of the TMD
operator contains additional divergences, which in general also depend
on the order of the $\vprp{b} $-derivative being taken, i.e., on which
$\vprp{k} $-moment is considered. In the case of higher $\vprp{k}
$-moments, the TMD operator does not reduce to a local operator in the
$b\rightarrow 0$ limit, but becomes a Qiu-Sterman type
quark-gluon-quark operator.  In general, different renormalization
factors could arise depending on the $\vprp{k} $-moment, i.e., the
renormalization factors in Eqs.~\eqref{rela2} and~\eqref{rela3}, or
within the group of relations
Eqs.~\eqref{rela4},~\eqref{rela5},~\eqref{rela6} and~\eqref{rela8}
need not be the same for $b\rightarrow 0$.

Finally, whereas the above discussion is premised on a multiplicative
renormalization pattern, it is not guaranteed that such a pattern
continues to hold when continuum symmetries, such as rotational
symmetry and chiral symmetry, are broken by the lattice
formulation. Absence of these symmetries often gives rise to operator
mixing, under which the numerators and denominators of the TMD ratios
considered here become sums of several terms, destroying
multiplicativity and the cancellation of renormalization factors in
the ratios. An example of such mixing, at one-loop in lattice
perturbation theory, for bilocal operators separated by a
straight-link path has recently been reported in Ref.~\cite{Constantinou:2017sej}.

The numerical data discussed in Sec.~\ref{sec:res} below yield a
varied picture with respect to these diverse possibilities, including
close agreement, within the present level of statistical accuracy,
between the two lattice ensembles for the three ratios,
Eqs.~\eqref{eq:GSS},~\eqref{eq:BMS} and~\eqref{eq:transversity}, even
at small $b$. On the other hand, at short $b$, significant differences
exist in the ratio defining the generalized $g_{1T} $ worm-gear shift
in the TMD limit $|\eta | \rightarrow \infty $, Eq.~\eqref{eq:WGS}.
Surprisingly, as discussed in Sec.~\ref{sec:straight}, these
differences persist even at large $b$ for the straight-link path,
i.e., in the case $\eta =0$. As discussed in Sec.~\ref{sec:straight}, it is
very likely that the observed effect is due to the
mixing reported in Ref.~\cite{Constantinou:2017sej} between the axial and
tensor operators that are used to calculate the generalized worm-gear
shift and the transversity.  While the straight-link case is not
directly relevant for TMD observables, it bears on operators used in studies of
PDFs
\cite{Ji:2013dva,Lin:2014zya,Alexandrou:2015rja,Ishikawa:2016znu,Chen:2016fxx,Chen:2016utp,Chen:2017mzz,Alexandrou:2016jqi,Alexandrou:2017huk}
and challenges our understanding of the renormalization of quark
bilocal operators.

\section{Lattice calculational scheme}
\label{sec:latt}

In order to utilize Lattice QCD techniques for the evaluation of the
fundamental correlator, Eq.~\eqref{spacecorr}, it is necessary to
boost the problem to a Lorentz frame in which the vectors $b$ and $v$
in Eq.~\eqref{spacecorr} are purely spatial; Minkowski temporal
separations cannot be accommodated in the Euclidean lattice setup. For
this reason, it is crucial to employ a definition of TMDs in which all
separations are space-like, cf.~the discussion in conjunction with
Eq.~\eqref{zetahatdef} above. With both $b$ and $v$ space-like, there
is no obstacle to the aforementioned boost. In addition, the
decomposition given in Eqs.~\eqref{atilde1}--\eqref{atilde3} of the
correlator into the invariant amplitudes $\widetilde{A}_{iB} $
facilitates translating the obtained data back into the original
Lorentz frame, i.e., results for observables cast in terms of these
amplitudes in the boosted frame are immediately valid also in the
latter.

%
%
%
%
%

\begin{table}[bp]
\begin{ruledtabular}
\begin{tabular}{c|cc}
ID             & Clover                     & DWF              \\\hline
Fermion Type   & Clover                     & Domain-wall      \\
Geometry       & $32^3 \times 96$           & $32^3 \times 64$ \\
$a (\fm)$      & 0.11403(77)                & 0.0840(14)       \\
$m_\pi (\MeV)$ & 317(2)(2)                  & 297(5)           \\
\# confs.      & 967                        & 533              \\
\# meas.       & 23208                      & 4264             \\
%
%
%
\end{tabular}
\end{ruledtabular}
\caption{Lattice parameters of the $n_f = 2+1$ flavor domain-wall
  ensemble generated by the RBC/UKQCD collaboration and the clover
  ensemble generated by the JLab/W\&M collaboration. The lattice
  spacings $a$ and pion masses $m_\pi$ for the clover and the DWF
  ensembles are quoted from Refs.~\cite{Green:2017keo}
  and~\cite{Syritsyn:2009mx}, respectively. Note that a different
  estimate of $a=0.127(1)$~fm is reported in~\cite{Yoon:2016jzj} for
  the clover ensemble, set using the Wilson flow parameter $w_0$,
  indicating that discretization errors are significant on these
  coarse lattices. }
\label{tab:ens}
\end{table}

The lattice parameters of the Wilson-clover and domain wall fermion
(DWF) ensembles analyzed in this work are summarized in
Table~\ref{tab:ens}.  The two ensembles have roughly the same pion
mass, about $300\MeV$, but differ in the lattice spacing. In
contrast to the previous study presented in Ref.~\cite{Musch:2011er},
we use unitary combinations of sea and valence quarks, \emph{i.e.}, we
use the same fermion discretization scheme for the sea and valence
quarks and the same values of the sea and valence quark masses.

\begin{table}
\centering
\begin{tabular}{c|l|c}
\hline\hline
$\vb/a$           & $\eta \vect{v}/a$ & $\vect{P}\cdot a L/(2\pi)$\\
\hline
$n\cdot(0,0,1)$   & $\pm n'\cdot(1,0,0)$  & $(0,0,0)$ , $(-1,0,0)$ \\
\cline{2-3}
$n = -7,\ldots,7$ (clover) & $\pm n'\cdot(1,1,0)$  & $(-1,0,0)$ \\
$n = -9,\ldots,9$ (DWF) & $\pm n'\cdot(1,-1,0)$ &            \\
\hline
$n\cdot(0,1,0)$   & $\pm n'\cdot(1,0,0)$  & $(0,0,0)$ , $(-1,0,0)$ \\
\cline{2-3}
$n = -7,\ldots,7$ (clover) & $\pm n'\cdot(0,0,1)$  & $(-1,0,0)$ \\
$n = -9,\ldots,9$ (DWF) & $\pm n'\cdot(1,0,1)$  &            \\
                  & $\pm n'\cdot(1,0,-1)$ &            \\
\hline
$n\cdot(0,1,1)$   & $\pm n'\cdot(1,0,0)$  & $(-1,0,0)$ \\
$n = -4,\ldots,4$ &                       &            \\
\hline
$n\cdot(0,1,-1)$  & $\pm n'\cdot(1,0,0)$  & $(-1,0,0)$ \\
$n = -4,\ldots,4$ &                       &            \\
\hline\hline
\end{tabular}
\caption{The parameters of the staple-shaped gauge connection,
  characterized by $\vb$ and $\eta \vv$, and the nucleon momenta
  $\vect{P}$ used in both the clover and the domain wall fermion
  calculations. $L$ is the spatial size of the lattice and for each $n$,
  the range of integers $n'$ is chosen to be large enough to cover the
  values for which one obtains a useful signal in the 3-point correlation
  functions. For example, in the clover fermion zero momentum case,
  $n^{\prime }_{\text{max}}$ is 15 for $|\vprp{b}|=0.11\fm$, 12 for
  $|\vprp{b}|=0.23\fm$, 11 for $|\vprp{b}|=0.34\fm$ and 10 for
  $|\vprp{b}|=0.45\fm$. In the case of off-axis Wilson lines, the
  TMD operator was improved by averaging over lattice paths approximating
  the continuum one; e.g., for $\vb = 2({\bf e}_{2} + {\bf e}_{3} )$,
  where ${\bf e}_{i} $ denotes the lattice link vector in $i$-direction,
  data were generated for both the sequence of links
  $({\bf e}_{2} ,{\bf e}_{3} ,{\bf e}_{2} ,{\bf e}_{3} )$ and the sequence
  of links $({\bf e}_{3} ,{\bf e}_{2} ,{\bf e}_{3} ,{\bf e}_{2} )$.
  % ...
\label{tab:meas_params}}
\end{table}

To describe the SIDIS and DY processes, we vary the nucleon three-momentum
$\vect{P}$, the separation $\vb$, the staple direction $\vv$ and the
corresponding length $\eta$ of the staple
as specified for both the clover and the DWF ensemble in
Table~\ref{tab:meas_params}. The resulting maximum magnitude of the
Collins-Soper parameter $\zetahat$ in this study is $|\zetahat|=0.32$
for the clover ensemble and $|\zetahat| = 0.41$ for the DWF ensemble.

In this study, we work in the isospin symmetric limit and calculate
matrix elements of only the isovector combination ($u-d$) of operators. In
this combination, the contributions of the disconnected quark loop
diagrams cancel.

\section{Numerical Results}
\label{sec:res}

Results for four TMD observables are presented in this section: the T-odd
Sivers and Boer-Mulders shifts, as well as the T-even generalized
transversity and worm-gear shift associated with the worm-gear TMD
$g_{1T} $. A more detailed description and physical interpretation of
these observables is given in Ref.~\cite{Musch:2011er}.

\subsection{The Generalized Sivers Shift}
\label{sec:Sivers}
The generalized Sivers shift addresses the distribution of transverse
momentum of unpolarized quarks in a transversely polarized nucleon,
where the transverse momentum direction and the nucleon polarization
direction are orthogonal to one another. It is defined by
\begin{align}
  \langle \vect{k}_y \rangle_{TU}(\vprp{\bvec}^2;\ldots) 
  \equiv \mN \frac{\tilde f_{1T}^{\perp[1](1)}(\vprp{\bvec}^2;\ldots)}
  {\tilde f_1^{[1](0)}(\vprp{\bvec}^2;\ldots)} \,,
\label{eq:GSS}
\end{align}
where $\mN$ is the nucleon mass, $f_{1T}^{\perp}$ is the Sivers TMD
\cite{Sivers:1989cc}, and $f_1$ is the unpolarized TMD. In
constructing this ratio, we use unrenormalized operators,
cf.~Eqs.~\eqref{atilde1},~\eqref{rela1} and~\eqref{rela7}.  Our
assumption of the cancellation of the renormalization factors is
discussed in Sec.~\ref{sec:setup}.

\begin{figure*}[tb]
  \includegraphics[width=0.48\linewidth]{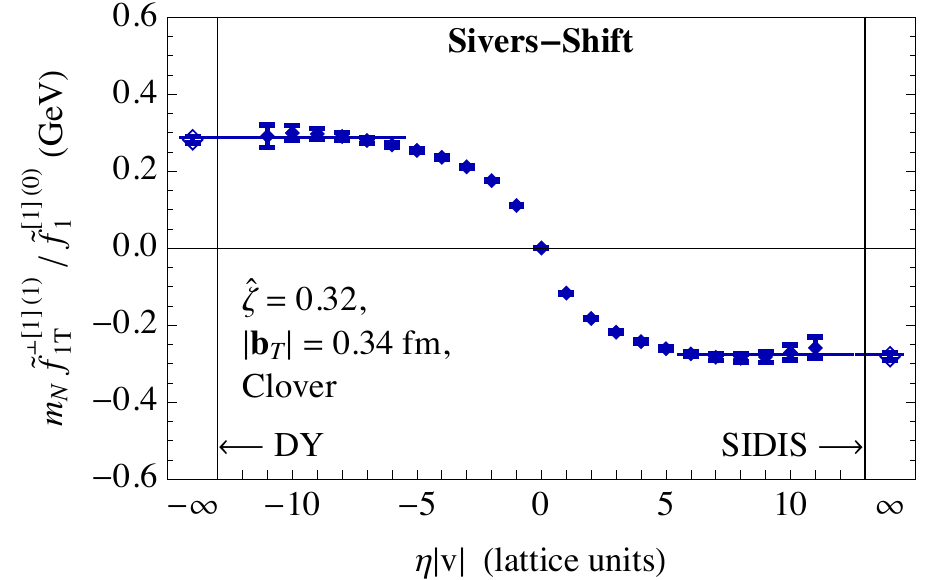}
  \includegraphics[width=0.48\linewidth]{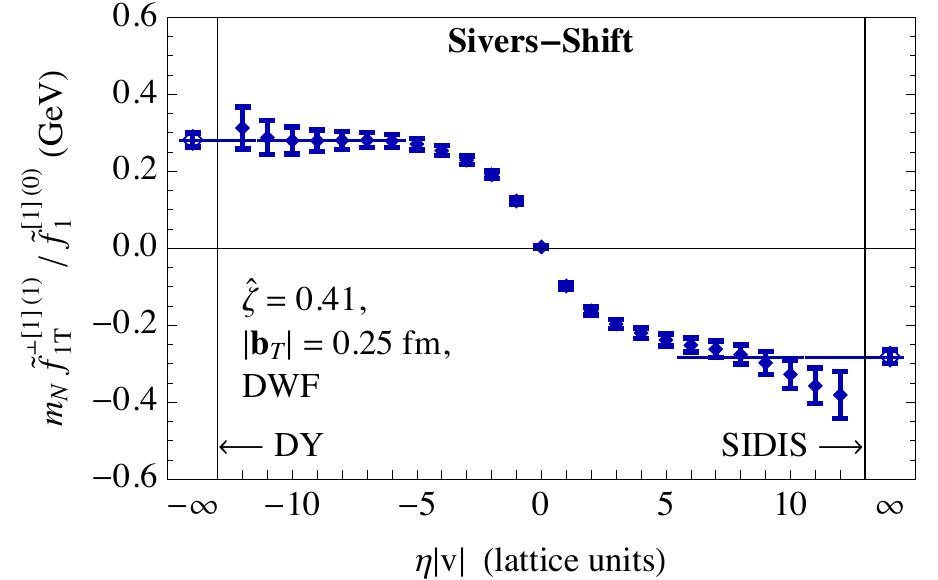}\\\vspace{1em}
  \includegraphics[width=0.48\linewidth]{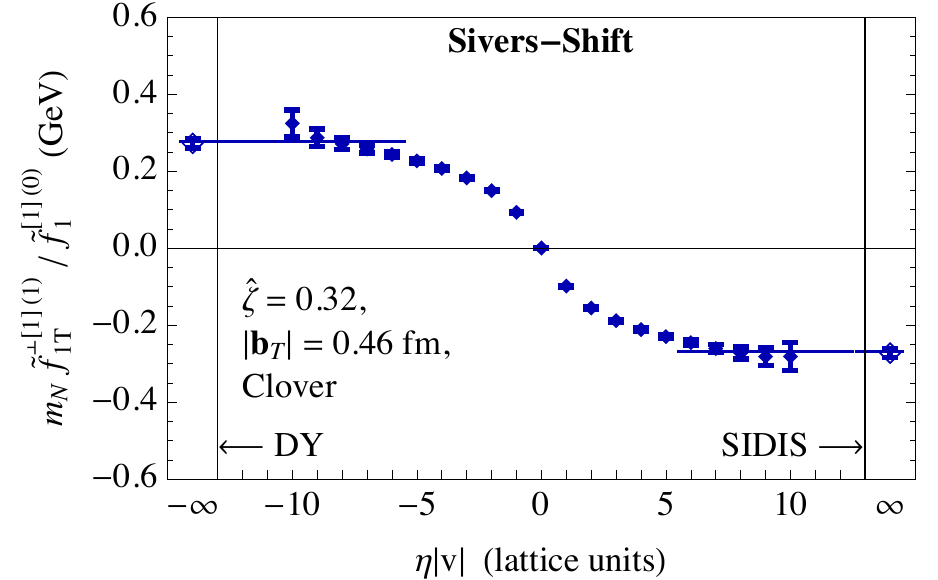}
  \includegraphics[width=0.48\linewidth]{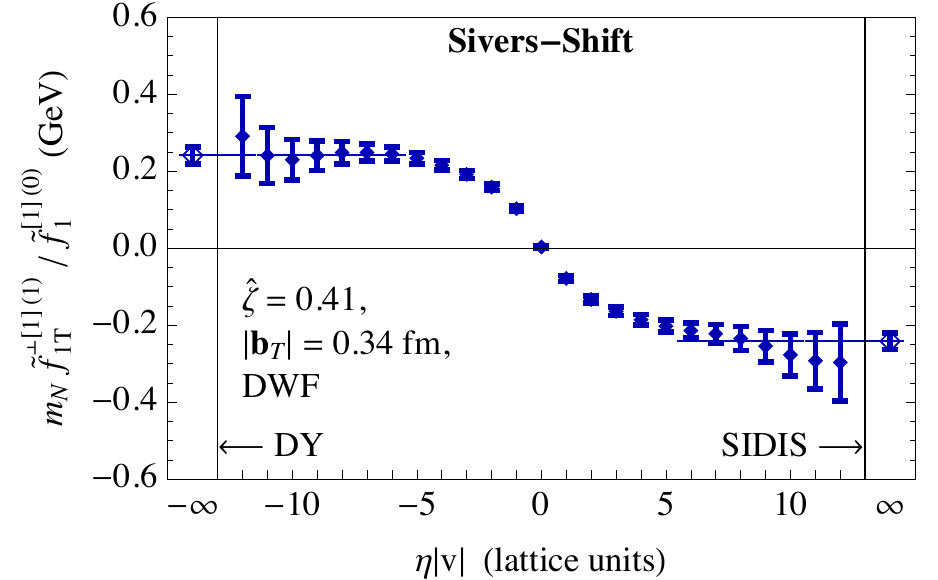}
\caption{
  Dependence of the generalized Sivers shift on the staple extent $\eta|\vv|$ 
  for the clover (left) and the DWF (right) ensembles at $|\vprp{b}|=3a$
  (top), and $4a$ (bottom). The Collins-Soper parameter is fixed at the
  highest value for which data are available, $\zetahat=0.41$ and 
  $0.32$ for the DWF and the clover ensembles, respectively. The asymptotic estimate is obtained 
  using a constant fit to data with $|\eta| \ge 6$. 
}
\label{fig:sivers-etav}
\end{figure*}

The dependence of the generalized Sivers shift on $\eta|\vv|$ is shown
in Fig.~\ref{fig:sivers-etav}.  In order to describe the SIDIS or the
DY process, the length of the staple $|\eta||\vv|$ connecting the
quark bilocal operator needs to be extrapolated to
infinity. In our setup, the DY process is obtained in the limit $\eta|\vv|
\rightarrow -\infty$, and SIDIS in the limit $\eta|\vv| \rightarrow
\infty$.

The data in Fig.~\ref{fig:sivers-etav} indicate
the onset of a plateau for $|\eta||\vv| \ge 6a$. A more stringent
estimate of the plateau value is obtained in the clover case that has
roughly a factor of six
larger statistics. Exploiting the evident T-odd behavior, we appropriately
average the data for $\pm |\eta||\vv| $ and fit them to a constant for
$|\eta||\vv| \ge 6a$.  This choice is based on the observation that the
results of fits starting at $|\eta||\vv| = 5a,\ 6a$ and $7a$ are
consistent within $1\sigma$. We do not use weighted fits since the
points at smaller $|\eta||\vv|$ have smaller statistical errors but
larger unquantified systematic errors.

These fits give the magnitude of the results for both DY and SIDIS
processes and the sign is taken from the data shown in
Fig.~\ref{fig:sivers-etav}. The error estimates are obtained using a
jackknife method. We find that the statistical uncertainties in the data
increase with both $\eta|\vv|$ and $|\vprp{b}|$.

\begin{figure*}[tb]
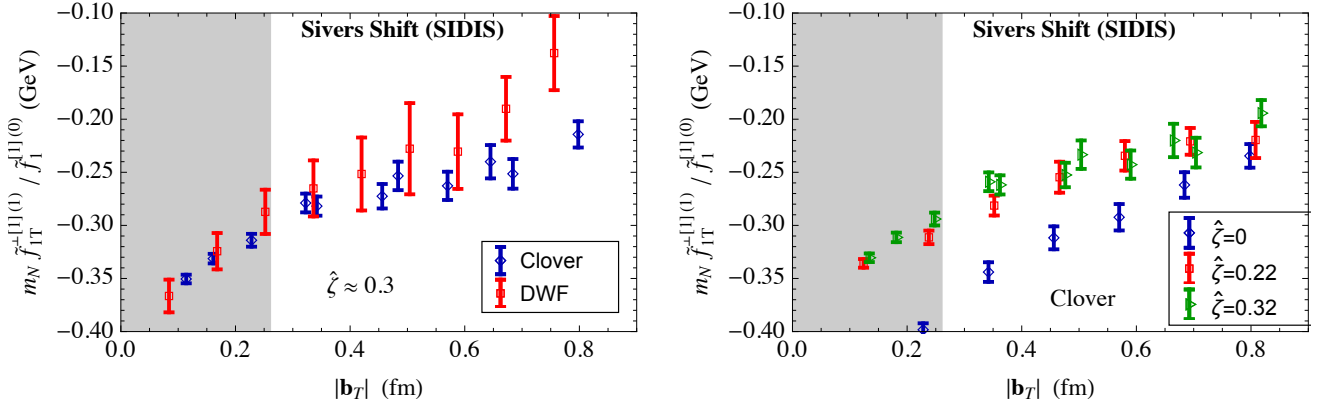

  \includegraphics[width=0.48\linewidth]{{{figs/UminusD_SiversRat_zetahat-0.35_bdepend_comb}}}\quad
  \includegraphics[width=0.48\linewidth]{{{figs/UminusD_SiversRat_zetahat_bdepend_Clover}}}
\caption{ Dependence of the generalized Sivers shift on
  $|\vprp{b}|$. In the left panel we compare DWF and clover results
  for $\zetahat \approx 0.3$ and in the right panel we show the higher
  precision clover data for three values of $\zetahat$.  The shaded
  area, $|\vprp{b}| \le 3a_{\text{DWF}} \approx 0.25\fm$, marks the
  region in which discretization effects could be expected.  }
\label{fig:sivers-evol}
\end{figure*}
\begin{figure*}[tb]
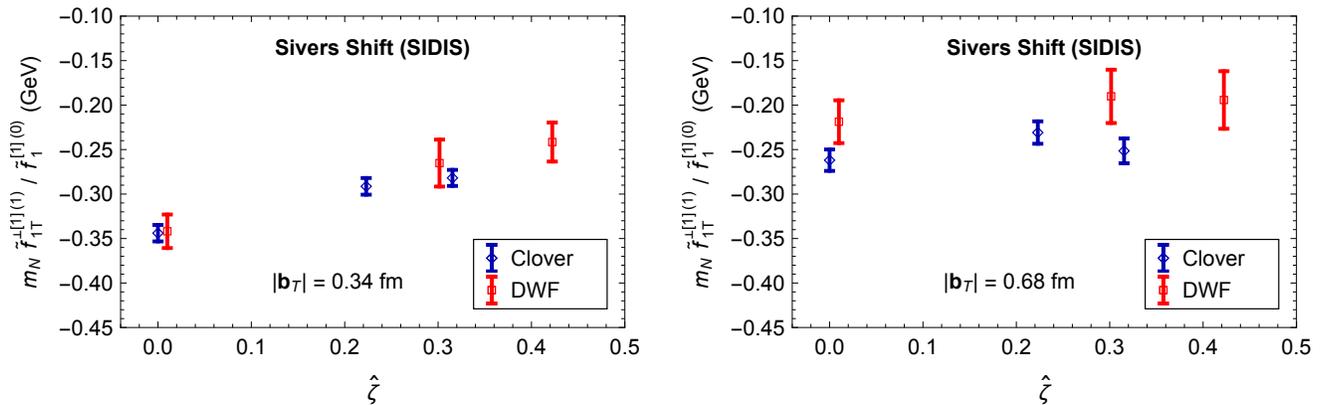

  \includegraphics[width=0.48\linewidth]{{{figs/UminusD_SiversRat_bT-0.34_evolution_comb}}}\quad
  \includegraphics[width=0.48\linewidth]{{{figs/UminusD_SiversRat_bT-0.68_evolution_comb}}}
\caption{
  Dependence of the generalized Sivers shift on $\zetahat$ for the two different ensembles and 
  for two values of $|\vprp{b}| =0.34$~fm (left) and 0.68~fm (right). 
}
\label{fig:sivers-evolX}
\end{figure*}

The dependence of the generalized Sivers shift on $|\vprp{b}|$ in the
SIDIS limit is compared for the two different ensembles in
Fig.~\ref{fig:sivers-evol} and the dependence on $\zetahat$ in
Fig.~\ref{fig:sivers-evolX}. The small $|\vprp{b}|$ region,
$|\vprp{b}| \le 3a_{\text{DWF}} \approx 0.25\fm$, in which lattice
artifacts and incomplete cancellation of renormalization factors could
be expected, is highlighted by the shaded region in
Fig.~\ref{fig:sivers-evol}, and also in Figs.~\ref{fig:boer-evol},
\ref{fig:h1-evol}, \ref{fig:g1t-evol}, and \ref{fig:straight} for the
other TMD observables. For larger $|\vprp{b}|$, the data in the SIDIS and DY
limit from the clover and the DWF ensembles are consistent for all
four TMDs analyzed in this work, suggesting that the fermion
discretization scheme and the lattice spacing effects are small. An
exception to this pattern is found, however, for the straight-link
path case, $\eta =0$, discussed in Sec.~\ref{sec:straight}.

The data for the Sivers shift in
Fig.~\ref{fig:sivers-evol}, and also the Boer-Mulders shift in
Fig.~\ref{fig:boer-evol}, start to show about $2\sigma$ deviation
between the two ensembles for $|\vprp{b}| > 0.6\fm$. At this
separation, the statistical errors in the DWF data are large and we
ascribe these deviations to statistical fluctuations; it should be
noted that, at these separations, contractions at large $|\eta |$,
which cease to provide a useful signal, are not evaluated and
therefore do not enter the plateau fits. This may also lead to an
underestimate of the uncertainty of the plateau value. On the other
hand, for the Sivers shift as well as the Boer-Mulders shift discussed
in Sec.~\ref{sec:Boer}, the agreement between the DWF and clover data persists into the
region of small $|\vprp{b}|$. Our conclusion is that, within errors,
no significant differences are seen between data from the two
ensembles for these two T-odd TMD observables, even in the limit of
small $\vprp{b} $.


%

In Fig.~\ref{fig:sivers-evolX}, we show the dependence of the
generalized Sivers shift on the Collins-Soper evolution parameter
$\zetahat$ for fixed $|\vprp{b}|=0.34$ and 0.68~fm.  The data from the
two ensembles are consistent, with the data for $|\vprp{b}|=0.68$~fm
showing less dependence on $\zetahat$. Since we have estimates only up
to $\zetahat = 0.41$, a future goal is to extend the calculation of
the TMD observables to large enough $\zetahat$, from where they can be evolved to
the light-cone limit, $\zetahat\rightarrow\infty$, using perturbation
theory.
%

In Section~\ref{sec:comp-sivers}, we compare these lattice estimates
for the generalized Sivers shift with one extracted from experimental
data at $\zetahat=0.83$. While the trend in the lattice data with
$\zetahat < 0.4$ suggests agreement at $\zetahat \sim 0.8$, we
consider it important to obtain data for $0.4 < \zetahat < 0.8$ to
establish this connection.  To increase $\zetahat$, however, requires
simulations with larger nucleon momenta. A recently developed
method~\cite{Bali:2016lva} that controls the rapid growth in
statistical errors with momenta~\cite{Musch:2011er} is under
investigation.
%

\subsection{The Generalized Boer-Mulders Shift}
\label{sec:Boer}
The second T-odd TMD observable we evaluate is the generalized
Boer-Mulders shift defined by
\begin{align}
  \langle \vect{k}_y \rangle_{UT}(\vprp{\bvec}^2;\ldots) 
  \equiv \mN \frac{\tilde h_{1}^{\perp[1](1)}(\vprp{\bvec}^2;\ldots)}
  {\tilde f_1^{[1](0)}(\vprp{\bvec}^2;\ldots)} \,.
\label{eq:BMS}
\end{align}
The Boer-Mulders function $h_{1}^{\perp}$~\cite{Boer:1997nt} describes
the distribution of the transverse momentum of transversely polarized
quarks in an unpolarized hadron, where the quark transverse momentum
and polarization are orthogonal to one another. 

\begin{figure*}[tb]
  \includegraphics[width=0.48\linewidth]{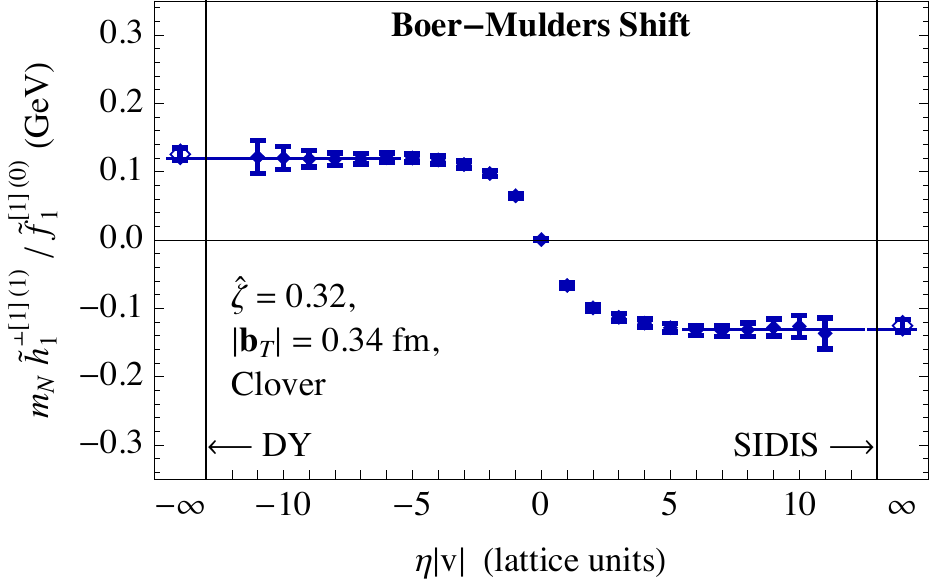}
  \includegraphics[width=0.48\linewidth]{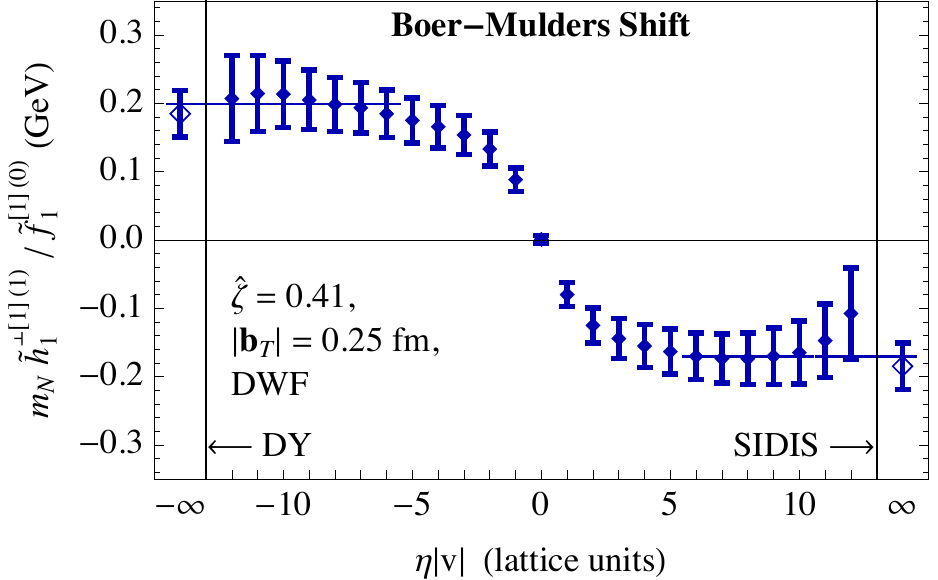}\\\vspace{1em}
  \includegraphics[width=0.48\linewidth]{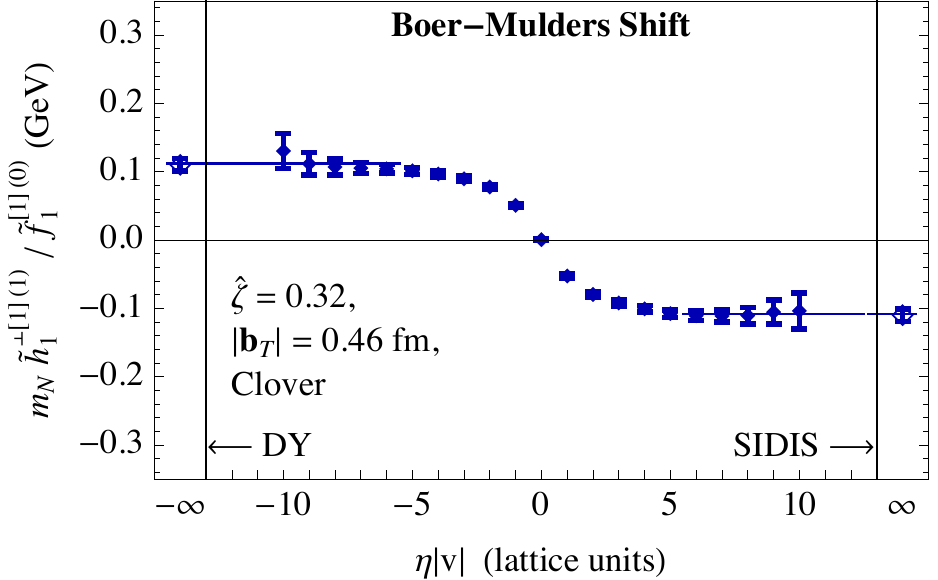}
  \includegraphics[width=0.48\linewidth]{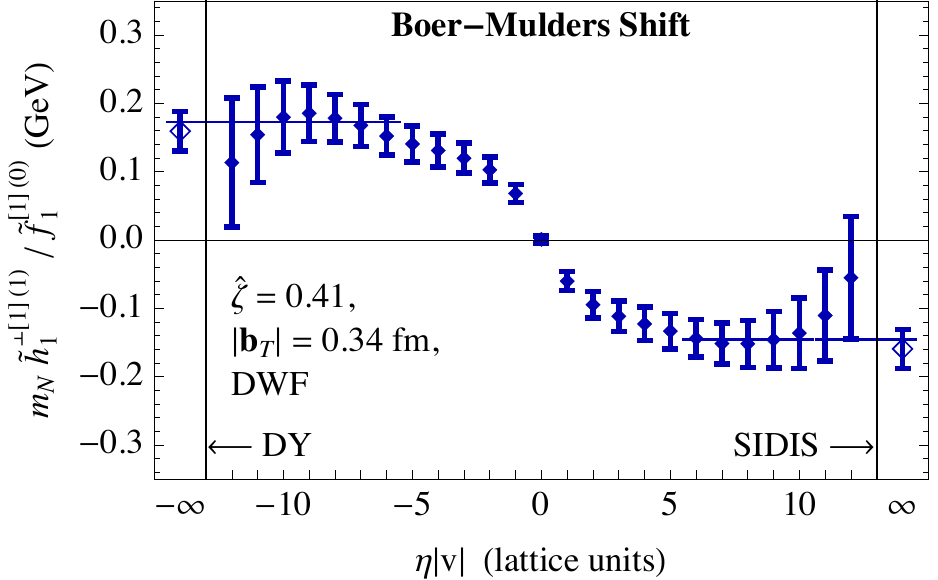}
\caption{ Dependence of the generalized Boer-Mulders shift on the
  staple extent $\eta|\vv|$ for the clover (left) and the DWF (right)
  ensembles. The rest is
  the same as in Fig.~\protect\ref{fig:sivers-etav}.  }
\label{fig:boer-etav}
\end{figure*}

The dependence of the generalized Boer-Mulders shift on $\eta|\vv|$ is
shown in Fig.~\ref{fig:boer-etav}. The data show a plateau at earlier
$|\eta||\vv|$ as compared to the Sivers shift; nevertheless, to preserve 
uniformity we again extrapolate to the DY and SIDIS limits using a constant
fit to data with $|\eta||\vv| \ge 6a$. Again these results are
consistent with those obtained with $|\eta||\vv| \ge 5a$ or $ 7a$.

\begin{figure*}[tb]
  \includegraphics[width=0.48\linewidth]{{{figs/UminusD_BoerMuldersRat_zetahat-0.35_bdepend_comb}}}\quad
  \includegraphics[width=0.48\linewidth]{{{figs/UminusD_BoerMuldersRat_zetahat_bdepend_Clover}}}
\caption{Dependence of the generalized Boer-Mulders shift 
  on $|\vprp{b}|$ for the two ensembles (left), and for three different
  values of $\zetahat$ analyzed on the clover ensemble (right).  The
  rest is the same as in Fig.~\protect\ref{fig:sivers-evol}.  }
\label{fig:boer-evol}
\end{figure*}
\begin{figure*}[tb]
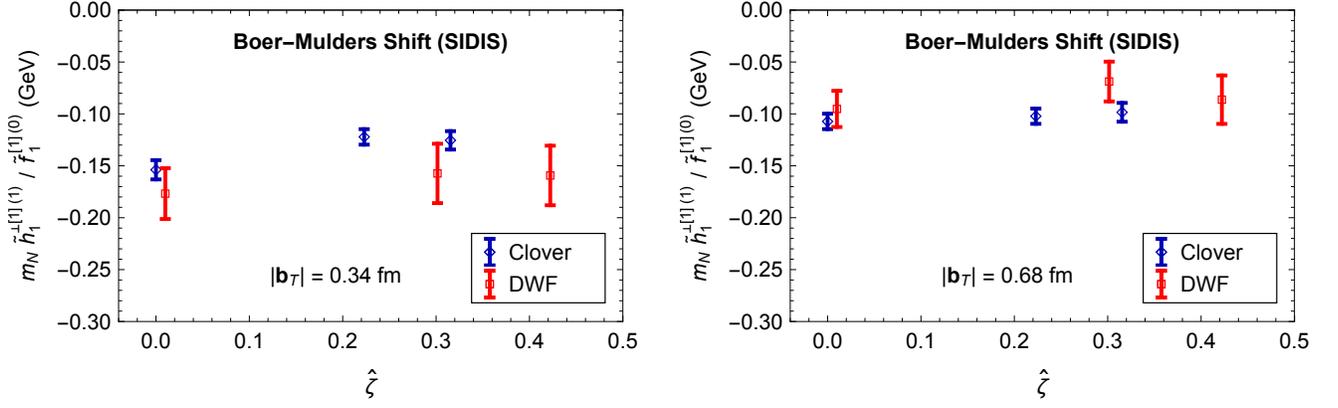

  \includegraphics[width=0.48\linewidth]{{{figs/UminusD_BoerMuldersRat_bT-0.34_evolution_comb}}}\quad
  \includegraphics[width=0.48\linewidth]{{{figs/UminusD_BoerMuldersRat_bT-0.68_evolution_comb}}}
\caption{Dependence of the generalized Boer-Mulders shift on
  $\zetahat$ for two values of $|\vprp{b}| =0.34$~fm (left) and
  0.68~fm (right).  }
\label{fig:boer-evolX}
\end{figure*}

The comparison of the dependence of the Boer-Mulders shift on
$|\vprp{b}|$ and $\zetahat$ between the clover and the DWF ensembles
is shown in Figs.~\ref{fig:boer-evol} and~~\ref{fig:boer-evolX}. We
again find that the results are compatible within their statistical
uncertainty over the entire range of $|\vprp{b}|$; no dependence on
the lattice action is observed even in the limit of small
$|\vprp{b}|$.  In Ref.~\cite{Engelhardt:2015xja}, the dependence of
the generalized Boer-Mulders shift on $\zetahat$ for pions was studied
up to $\zetahat=2.03$ by taking advantage of the lighter mass and
better signal-to-noise ratio in pion correlation functions as compared
to those for nucleons.  Results for the pion show that a significant
portion of the evolution to large $\zetahat $ is already achieved when
$\zetahat \sim 2$.

The higher statistics clover data in the right panel of
Figs.~\ref{fig:sivers-evol} and~\ref{fig:boer-evol} show that the two
T-odd TMD observables of the nucleon, the Sivers and the Boer-Mulders shifts
(SIDIS case), increase with $\zetahat$ and $|\vprp{b}|$, and the data
at the three values of $\zetahat$ have, within 1$\sigma$ errors,
converged by $|\vprp{b}| \approx 0.8$~fm. 

\subsection{The Transversity $h_1$}
The T-even TMDs, unlike the T-odd TMDs such as the Sivers and Boer-Mulders
distributions, are process-independent, \emph{i.e.}, the same for
DY and SIDIS processes. They were initially studied in Lattice QCD in
a truncated fashion by using a straight Wilson
line~\cite{Hagler:2009mb, Musch:2010ka} and the treatment was subsequently
extended to the physically relevant case of staple-shaped paths describing
the SIDIS and DY processes. It has been observed that the difference between
the two approaches is in many cases small for T-even TMDs \cite{Musch:2011er},
i.e., there is only a mild $\eta $-dependence. In this study, our
observations are similar for the two different lattice
discretization schemes, and at the lighter pion masses investigated,
although the picture for the $g_{1T} $ worm-gear shift discussed in
Secs.~\ref{sec:WGS} and \ref{sec:straight} is not as clear-cut.

The first T-even observable we present is the generalized tensor charge
defined by the ratio between transversity and the unpolarized function:
\begin{align}
  \frac{\tilde h_{1}^{[1](0)}(\vprp{\bvec}^2;\ldots)}
  {\tilde f_1^{[1](0)}(\vprp{\bvec}^2;\ldots)} \,.
\label{eq:transversity}
\end{align}
It is called the generalized tensor charge because the integral of the
transversity, obtained in position space by setting
$\vprp{b}^2\!=\!0$, formally gives the nucleon tensor charge: 
$g_T^{u-d}=\int\! dx\, d^2\vprp{k}\, h_1(x,\vprp{k}^2)=\tilde h_{1}^{[1](0)}(\vprp{b}^2\!=\!0)$.
\begin{figure*}[tb]
  \includegraphics[width=0.48\linewidth]{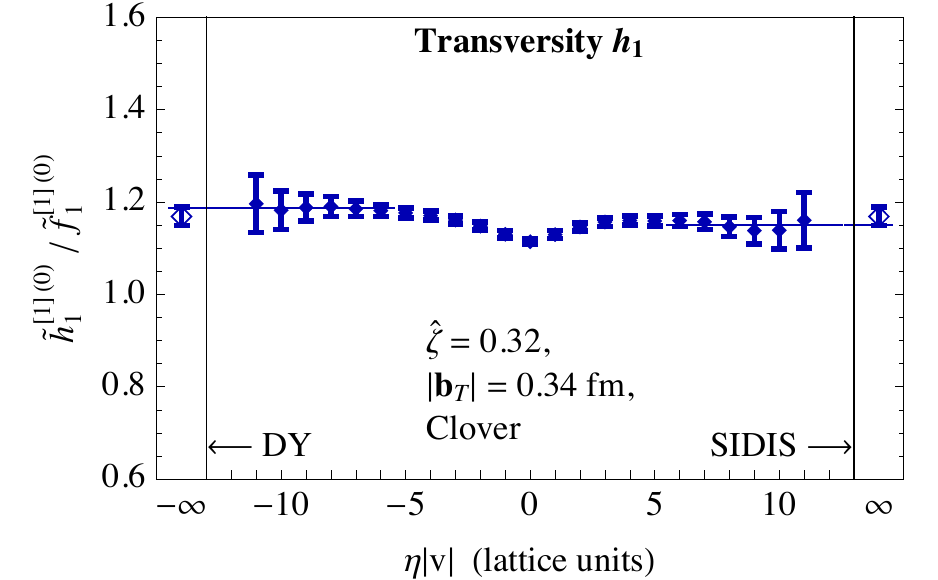}
  \includegraphics[width=0.48\linewidth]{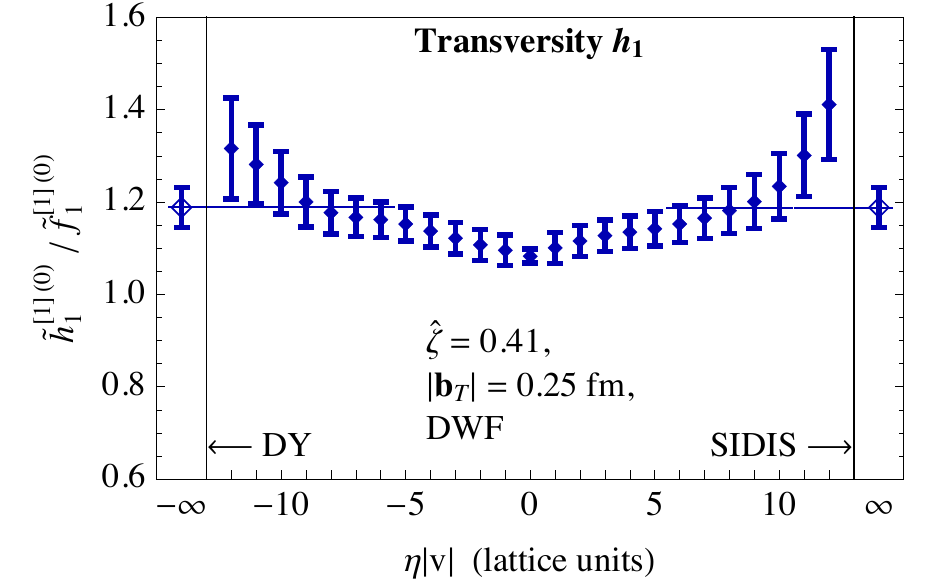}\\\vspace{1em}
  \includegraphics[width=0.48\linewidth]{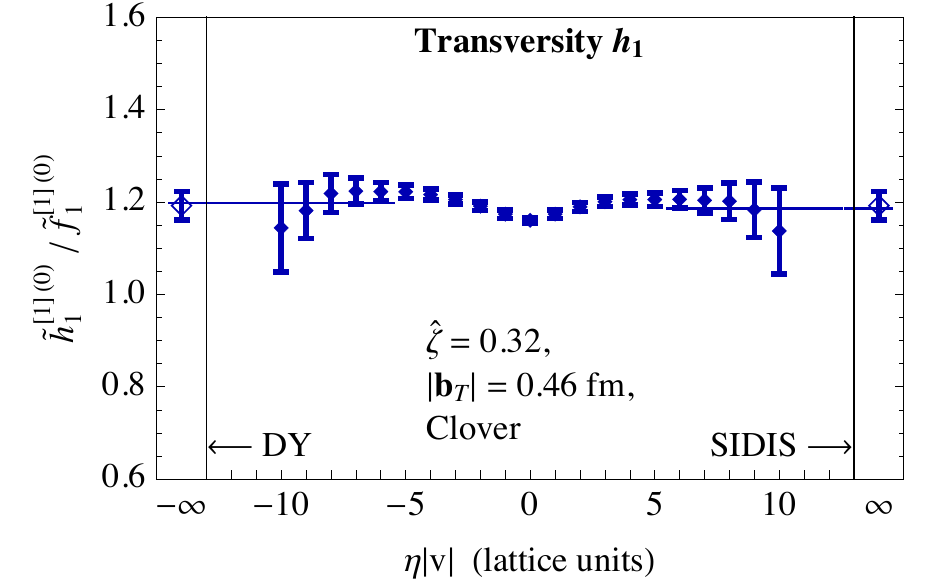}
  \includegraphics[width=0.48\linewidth]{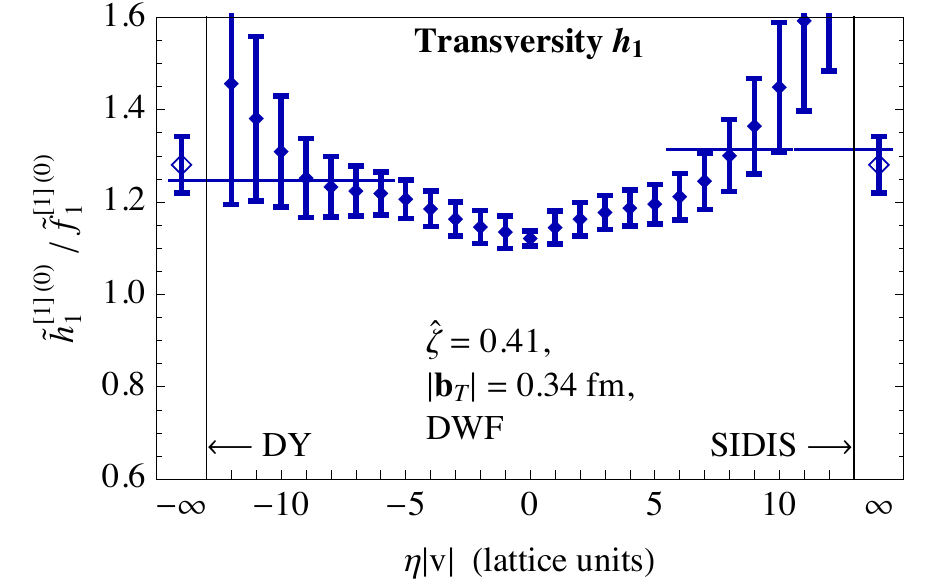}
\caption{Dependence of the  transversity ratio
  $\tilde h_{1}^{[1](0)}/\tilde f_1^{[1](0)}$ on the staple extent
  $\eta|\vv|$ for the clover (left) and the DWF (right) ensembles. 
  The rest is the same as in Fig.~\protect\ref{fig:sivers-etav}.
}
\label{fig:h1-etav}
\end{figure*}

\begin{figure*}[tb]
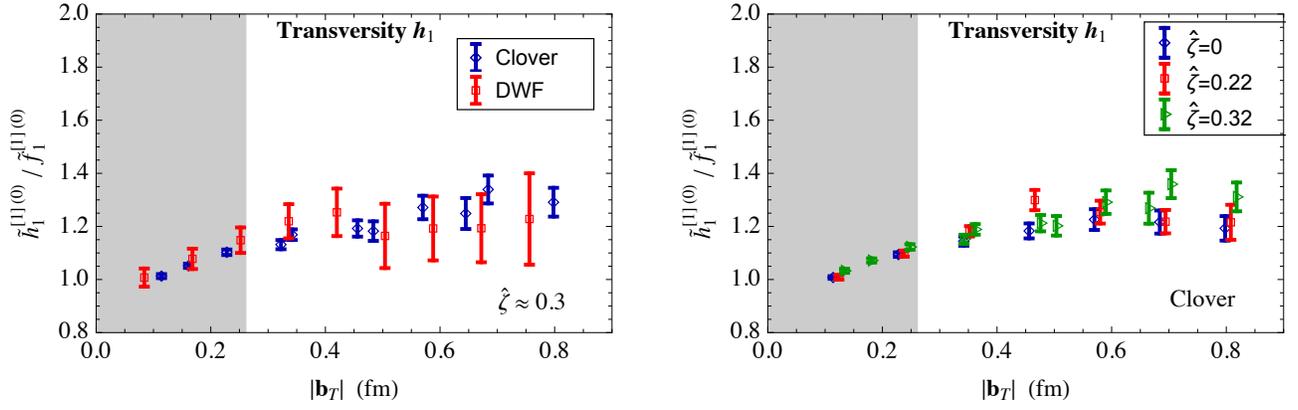

  \includegraphics[width=0.48\linewidth]{{{figs/UminusD_h1Rat_zetahat-0.35_bdepend_comb}}}\quad
  \includegraphics[width=0.48\linewidth]{{{figs/UminusD_h1Rat_zetahat_bdepend_Clover}}}
\caption{Dependence of the transversity ratio $\tilde h_{1}^{[1](0)}/\tilde f_1^{[1](0)}$
  on $|\vprp{b}|$ for the two ensembles (left), and for three different
  values of $\zetahat$ analyzed on the clover ensemble (right).  The
  rest is the same as in Fig.~\protect\ref{fig:sivers-evol}.  }
\label{fig:h1-evol}
\end{figure*}

\begin{figure*}[tb]
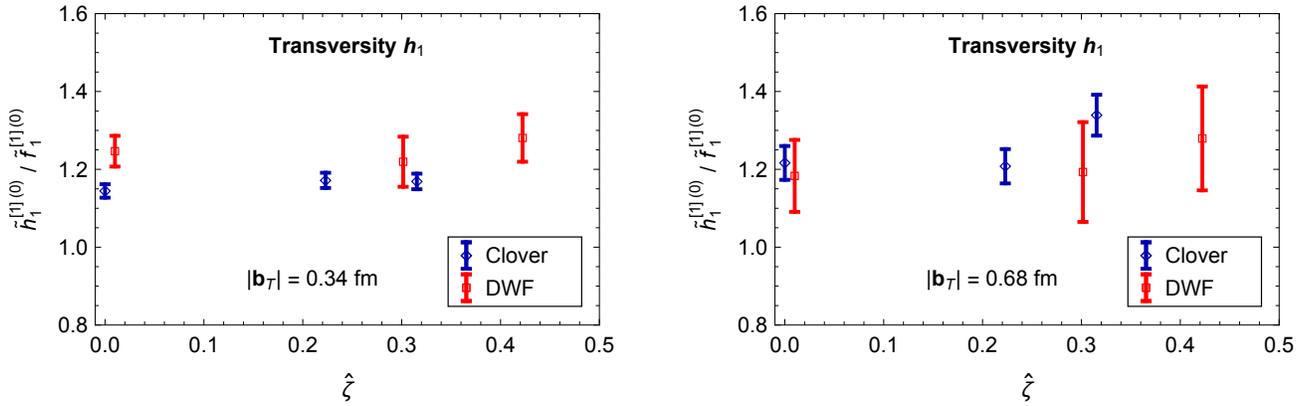

  \includegraphics[width=0.48\linewidth]{{{figs/UminusD_h1Rat_bT-0.34_evolution_comb}}} \quad
  \includegraphics[width=0.48\linewidth]{{{figs/UminusD_h1Rat_bT-0.68_evolution_comb}}} \quad
\caption{Dependence of the transversity ratio
  $\tilde h_{1}^{[1](0)}/\tilde f_1^{[1](0)}$
  on $\zetahat$ for two values of $|\vprp{b}| =0.34$~fm (left) and 0.68~fm (right).  }
\label{fig:h1-evolX}
\end{figure*}

The data for the transversity ratio given in Fig.~\ref{fig:h1-etav}
show that the $\eta|\vv|$-dependence is much smaller than for the T-odd
TMDs but non-zero. The DWF data are noisy and do not show a clear plateau. The
higher statistics clover data, and a previous study using a mixed-action
DWF-on-AsqTad lattice scheme at $m_\pi=518\MeV$ with
$\zetahat=0.39$ \cite{Musch:2011er}, show a plateau from which the
asymptotic value can be extracted.  We again fit the data with
$|\eta||\vv| \ge 6a$ to a constant for both ensembles.
The $|\vprp{b}|$ and $\zetahat$ dependences of the transversity ratio
are illustrated in Figs.~\ref{fig:h1-evol} and ~\ref{fig:h1-evolX}.  The data in the left panel of 
Fig.~\ref{fig:h1-evol} for 
both ensembles show a consistent plateau for $|\vprp{b}| > 0.3$~fm.
Again, as in the case of the T-odd TMDs, the DWF and the clover data
agree even in the regime of small $|\vprp{b}|$. Also, the data in Fig.~\ref{fig:h1-evolX}
show no significant dependence on $\zetahat$.
%

\subsection{The generalized $g_{1T}$ worm-gear shift}
\label{sec:WGS}
\begin{figure*}[tb]
  \includegraphics[width=0.48\linewidth]{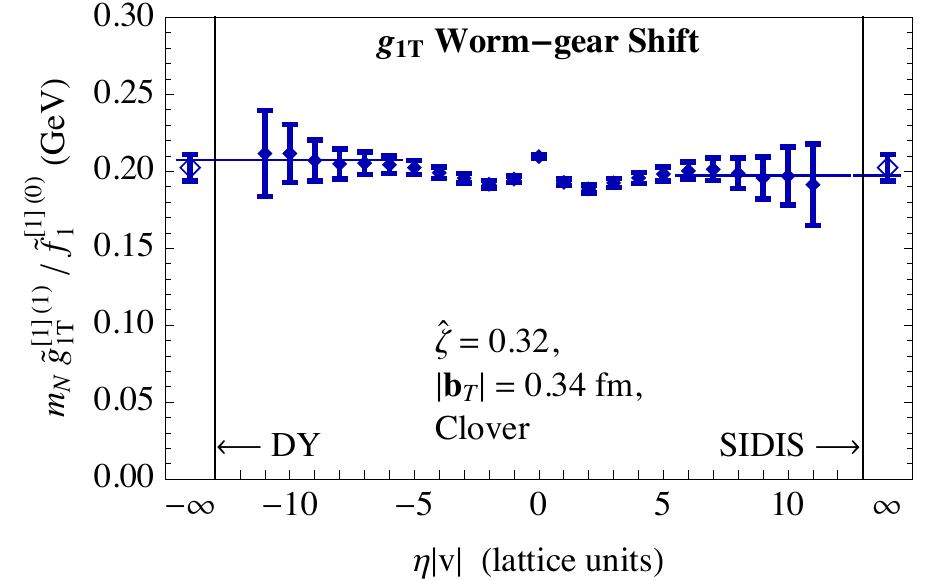}
  \includegraphics[width=0.48\linewidth]{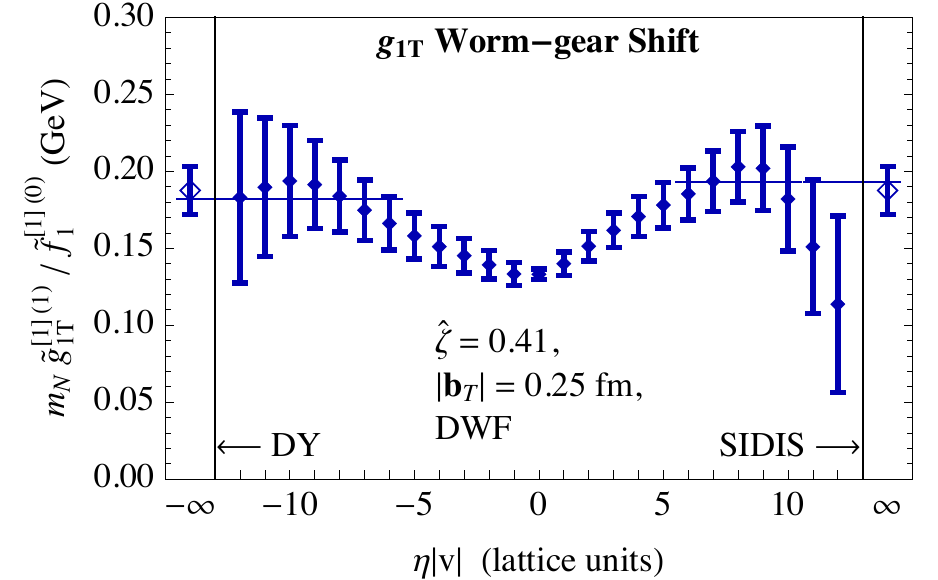}\\\vspace{1em}
  \includegraphics[width=0.48\linewidth]{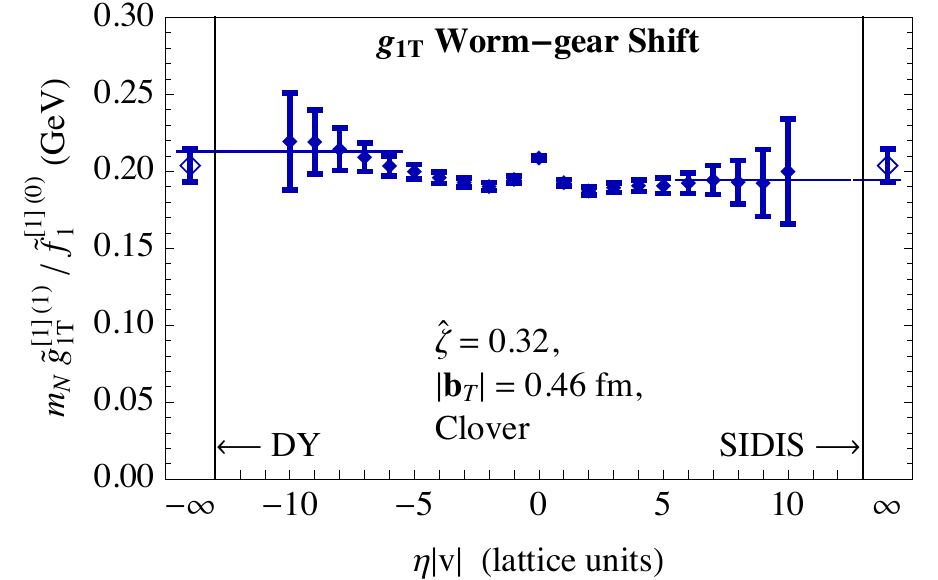}
  \includegraphics[width=0.48\linewidth]{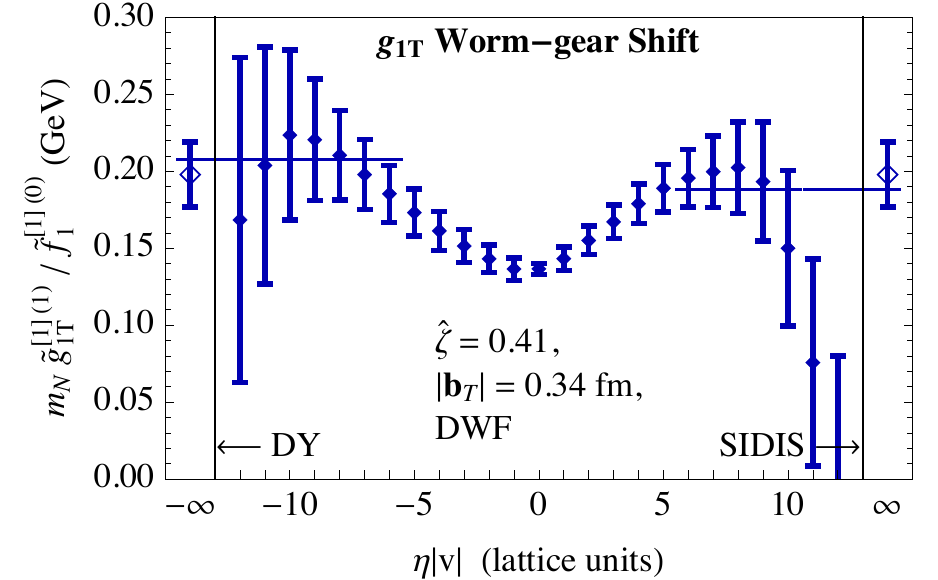}
\caption{
  Dependence of the generalized $g_{1T}$ worm-gear shift on the staple extent $\eta|\vv|$
  for the clover (left) and the DWF (right) ensembles.
  The rest is the same as in Fig.~\protect\ref{fig:sivers-etav}.
}
\label{fig:g1t-etav}
\end{figure*}

The second T-even TMD observable considered in this work is the generalized
worm-gear shift defined by
\begin{align}
  \langle \vect{k}_x \rangle_{TL}(\vprp{\bvec}^2;\ldots) 
  \equiv \mN \frac{\tilde g_{1T}^{[1](1)}(\vprp{\bvec}^2;\ldots)}
  {\tilde f_1^{[1](0)}(\vprp{\bvec}^2;\ldots)} \,,
\label{eq:WGS}
\end{align}
where $g_{1T}$ is one of the the ``worm-gear'' functions, the
transversal helicity \cite{Tangerman:1994eh}. The dependence of the
generalized $g_{1T}$ shift on $\eta|\vv|$ is shown in
Fig.~\ref{fig:g1t-etav}.  Similar to what is observed in the
transversity ratio, the generalized $g_{1T}$ shift on the clover
lattices shows little change in the transition from the straight
Wilson line ($\eta = 0$) to the staple-shaped path, other than the
cusp at $\eta = 0$. The DWF data show a dependence on $\eta$, but
note that the uncertainties are large. Fig.~\ref{fig:g1t-evol} shows the
dependence of the SIDIS (or equivalently DY) limits of the generalized $g_{1T}$ shift
on $|\vprp{b}|$ and
$\zetahat$ for the two different ensembles. Again, the results from
the two ensembles are consistent, as expected, for $|\vprp{b}| \ge
0.3$~fm. 

\begin{figure*}[tb]
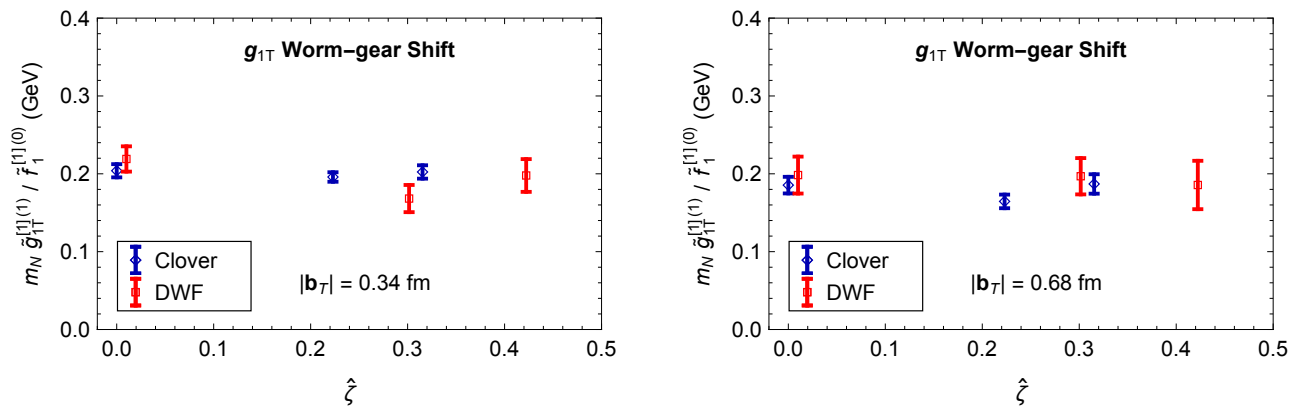

  \includegraphics[width=0.48\linewidth]{{{figs/UminusD_g1TRat_bT-0.34_evolution_comb}}}\quad
  \includegraphics[width=0.48\linewidth]{{{figs/UminusD_g1TRat_bT-0.68_evolution_comb}}}
\caption{ Dependence of the generalized $g_{1T}$ worm-gear shift on
  $\zetahat$ for two values of
  $|\vprp{b}| =0.34$~fm (left) and 0.68~fm (right).  }
\label{fig:g1t-evolX}
\end{figure*}

Both the worm-gear shift (Fig.~\ref{fig:g1t-evolX}) 
and the transversity (Fig.~\ref{fig:h1-evolX}) 
show little dependence on $\zetahat$ in
contrast to the data for the T-odd shifts given in
Fig.~\ref{fig:sivers-evol} and~\ref{fig:boer-evol} which 
show a significant difference between the $\zetahat=0$ and
$\zetahat=0.22$ or $0.32$ cases, especially at small $|\vprp{b}|$.

The generalized $g_{1T}$ worm-gear shift does differ qualitatively 
from the other TMD ratios studied, in that a
significant difference between the two ensembles is observed when $|\vprp{b}| \le 0.25\fm$. 
Further data are needed to clarify whether this
difference is due to the failure of the cancellation of
renormalization factors in the ratios as discussed in
Sec.~\ref{sec:setup} or different discretization effects in the two
lattice formulations. It is important to bear in mind that for the TMD
observable of interest, the generalized $g_{1T}$ worm-gear shift, the two lattice
formulations give consistent results for $|\vprp{b}| \ge 0.3$~fm as
shown in Fig.~\ref{fig:g1t-evol}.

\begin{figure*}[tb]
  \includegraphics[width=0.48\linewidth]{{{figs/UminusD_g1TRat_zetahat-0.35_bdepend_comb}}}\quad
  \includegraphics[width=0.48\linewidth]{{{figs/UminusD_g1TRat_zetahat_bdepend_Clover}}}
\caption{Dependence of the generalized $g_{1T}$ worm-gear shift 
  on $|\vprp{b}|$ for the two ensembles (left), and for three different
  values of $\zetahat$ analyzed on the clover ensemble (right).  The
  rest is the same as in Fig.~\protect\ref{fig:sivers-evol}.  }
\label{fig:g1t-evol}
\end{figure*}

Overall, in the SIDIS and DY limit, the data presented exhibit
consistency between the two lattice ensembles for all four observables
considered once the quark separation $|\vprp{b}|$ in the bilocal TMD
operator exceeds about three lattice spacings, indicating that, in the
regime of finite physical extent, the lattice operators approximate
the expected continuum behavior. Only in the case of the generalized
$g_{1T}$ worm-gear shift, significant differences between domain wall
and clover fermions are observed at small $|\vprp{b}|$. For the other
three observables, it is encouraging to note that the agreement
persists into the quasi-local regime.

\subsection{Transversity and worm-gear shift from straight gauge link paths}
\label{sec:straight}

To obtain further insight into the discrepancy between the data, at
small $\vprp{b} $, from the two lattice formulations in the
generalized $g_{1T} $ worm-gear shift and buttressed by the superior
statistical accuracy of the data when $\eta =0$, we examined also the
case of a straight gauge connection for the T-even TMD operators. It
should be emphasized that this is not the physically relevant case for
TMD studies; both the SIDIS and the DY processes are described by a
staple-shaped gauge connection with $|\eta| \to \infty$ that encodes
final and initial state interactions, respectively.  However, such
straight-link operators are used, e.g., in the study of PDFs in the
approach developed in
Refs.~\cite{Ji:2013dva,Lin:2014zya,Alexandrou:2015rja,Ishikawa:2016znu,Chen:2016fxx,Chen:2016utp,Chen:2017mzz,Alexandrou:2016jqi,Alexandrou:2017huk}.

\begin{figure*}[tb]
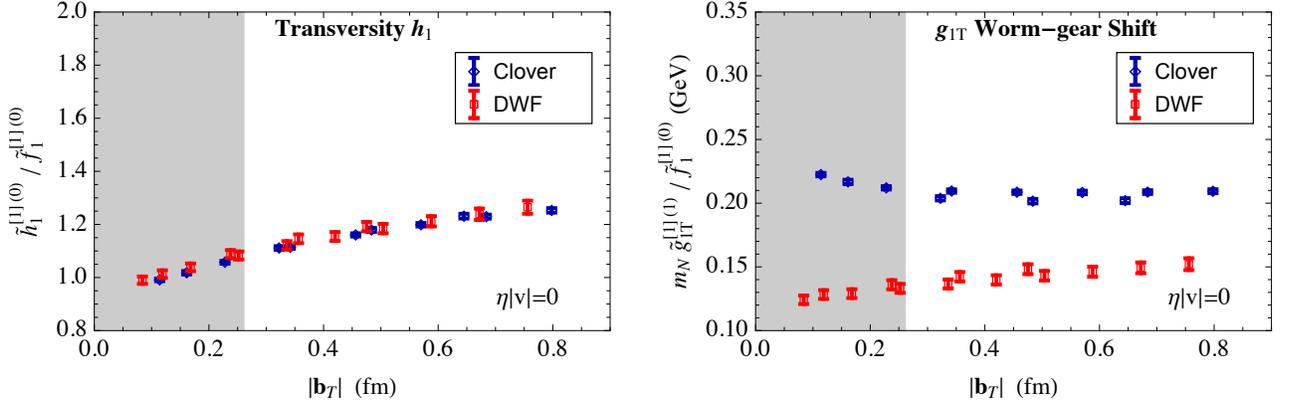

  \includegraphics[width=0.48\linewidth]{{{figs/UminusD_h1Rat_etahat-0.0_bdepend_comb}}}
  \includegraphics[width=0.48\linewidth]{{{figs/UminusD_g1TRat_etahat-0.0_bdepend_comb}}}
\caption{Dependence of the transversity (left) and generalized $g_{1T}
  $ worm-gear shift (right) on the length of the straight-link paths,
  $|\vprp{b}|$, for the two different ensembles. The striking
  observation is that the difference between the DWF and clover data
  for the worm-gear shift persists for all $|\vprp{b}|$. The data
  shown are for nucleon momentum $|\vect{P} |=2\pi/(aL)$; results
  for $\vect{P} =0$ coincide with these data within the uncertainties shown. }
\label{fig:straight}
\end{figure*}

The data for the two T-even quantities, the generalized worm-gear
shift and the transversity, for straight-link paths connecting the
quark fields are shown in Fig.~\ref{fig:straight} (the corresponding
data for the T-odd Sivers and Boer-Mulders shifts are consistent with
zero, as expected).  The data for the clover and DWF fermions agree
for the transversity $ \tilde h_{1}^{[1](0)}(\vprp{\bvec}^2;\ldots)/
{\tilde f_1^{[1](0)}(\vprp{\bvec}^2;\ldots)} $ for all values of
$\vprp{b} $ starting at a separation of one link, even at the improved
level of accuracy afforded by the straight-link case. This is
consistent with the pattern seen in Fig.~\ref{fig:h1-evol}. However,
examining the generalized worm-gear shift, $ \mN \tilde
g_{1T}^{[1](1)}(\vprp{\bvec}^2;\ldots)/{\tilde
  f_1^{[1](0)}(\vprp{\bvec}^2;\ldots)}$, one is confronted with the
surprising result that the data on the two ensembles differ for all
$\vprp{b} $.  The discrepancy observed in the staple-link case at only
small $\vprp{b} $ with $\eta \to \infty$, cf.~Fig.~\ref{fig:g1t-evol},
opens up for $\eta =0$ to persist for all $\vprp{b} $ considered. This
difference can be traced back to the opposite nature of the cusp in
the two data sets (DWF versus clover)  at $\eta =0$ as evident from
Fig.~\ref{fig:g1t-etav}.

The recent 1-loop lattice perturbation theory calculation, presented
in Ref.~\cite{Constantinou:2017sej}, shows that for clover fermions there
is a mixing of the straight-link bilocal axial and tensor quark
operators that we have used to calculate the generalized worm-gear
shift and the transversity. This mixing is a lattice artifact due to
the explicit breaking of the chiral symmetry in the clover
formulation.  To analyze the impact of the mixing on the clover data
in detail requires calculating the contributions of all the non-zero
Lorentz invariants in the axial (tensor) channel (See Eqs. (19) and
(20) in Ref.~\cite{Musch:2011er}), which we have not done.  In our
data, an effect is only seen in the worm-gear shift, but not in the
generalized transversity. This would be the expected behavior in a
scenario where the worm-gear shift, after taking into account
kinematic factors, is much smaller than the generalized
transversity. We speculate this to be the reason for the mixing
effects being manifest in only the worm-gear shift.

To summarize, our key observation is that a difference between DWF and
clover results is observed only in the case where there is a mixing
between operators, calculated at 1-loop in
Ref.~\cite{Constantinou:2017sej}. Whether the mixing, analyzed at 1-loop,
is the explanation for the full non-perturbative effect seen remains
to be confirmed by future calculations.

%

\section{Comparison with Experimental Estimate of Generalized Sivers Shift}
\label{sec:comp-sivers}

In this section, we compare the Lattice QCD calculation of the
generalized Sivers shift defined in Eq.~\eqref{eq:GSS} with the result
extracted from SIDIS experimental data. 

At leading order in perturbation theory~\cite{Echevarria:2014xaa}, 
the unpolarized function and the Sivers function are written as
\begin{align}
\tilde f_{1,q}^{(0)} (x, b; Q) &=  f_{q}(x, Q),
\label{eq:pdf-expand}
\\
\tilde f_{1T,q}^{\perp (1)}(x, \vprp{\bvec}^2;Q,\ldots ) 
&= -\frac{1}{2 \mN}  T_{q,F}(x, x; Q)\,,
\label{eq:sivers-expand}
\end{align}
where $f_{q}(x, Q)$ is the collinear PDF,
and $T_{q,F}(x, x, Q)$ is the twist-3 Qiu-Sterman quark-gluon correlation
function.
The $x$-integral of the collinear PDF is the number of valence quarks
in a proton, so the denominator of Eq.~\eqref{eq:GSS} becomes $1$ for
the $u-d$ isovector combination.
For the Qiu-Sterman function, Ref.~\cite{Echevarria:2014xaa} uses the 
ansatz
\begin{align}
T_{q, F}&(x, x, Q;~ \alpha_q, \beta, N_q) \nonumber \\
&= N_q \frac{(\alpha_q+\beta)^{(\alpha_q+\beta)}}{\alpha_q^{\alpha_q} \beta^{\beta}}
x^{\alpha_q} (1-x)^{\beta} f_{q}(x, Q),
\label{eq:QSfunc}
\end{align}
and the parameters $N_q, \alpha_q$ and $\beta$ are determined by a global
fit to the Sivers asymmetry data in SIDIS experiments at HERMES,
COMPASS and Jefferson Lab.
Following Ref.~\cite{Echevarria:2014xaa}, we take the Qiu-Sterman function
expressed in terms of fit parameters with errors and ignore the smaller
uncertainties in the collinear PDF, given in Ref.~\cite{Gao:2013xoa}.
Using these parameterized functions, the error in the generalized Sivers 
shift is estimated by generating a bootstrap sample for the numerator 
using a normal distribution with mean and error given in 
Ref.~\cite{Echevarria:2014xaa}. We choose the momentum scale 
$Q=\sqrt{2.4}$~GeV, which is the typical momentum scale of the 
HERMES experiments, large enough to expect perturbation theory to be 
reliable (i.e., $Q \gg \Lambda_\text{QCD}$), and 
close to the scale of our lattice calculations $(Q \approx 1/a)$. 
TMDs also depend on the variable $\zeta$, and the authors in
Ref.~\cite{Echevarria:2014xaa} use $\zeta = Q$, which corresponds to
$\zetahat = \zeta/ 2 \mN =0.83$ in our calculation.

With these simplifications, the generalized Sivers shift is defined via the 
$x$-integrals of the TMDs over the range $[-1, 1]$, with the data at 
negative values of $x$ given by the antiquark distribution. 
Note that in the numerator in Eq.~\eqref{eq:GSS}, the quark and 
antiquark distributions are summed, whereas in
the denominator, the antiquark distribution is subtracted from the quark 
distribution \cite{Tangerman:1994eh}.
The desired phenomenological estimate of the generalized Sivers shift 
for the isovector operator is then given by
\begin{align}
  \langle \vect{k}_y \rangle_{TU} ^\text{SIDIS}
  = \mN \frac{\tilde f_{1T,u}^{\perp[1](1)} - \tilde f_{1T,d}^{\perp[1](1)}}
    {\tilde f_{1,u}^{[1](0)} - \tilde f_{1,d}^{[1](0)} } = -0.146(49) \,.
\label{eq:gen_siv_res}
\end{align}
Note that this ratio, calculated using the leading order expressions given in 
Eqs.~\eqref{eq:pdf-expand} and~\eqref{eq:sivers-expand}, is
independent of $\vprp{\bvec}^2$ and the momentum scale.
The scale dependence cancels in the ratio at leading order in perturbation
theory, and thus, any reasonable choice should have a small impact on
the generalized Sivers shift.

\begin{figure*}[tb]
\includegraphics[width=0.48\textwidth]{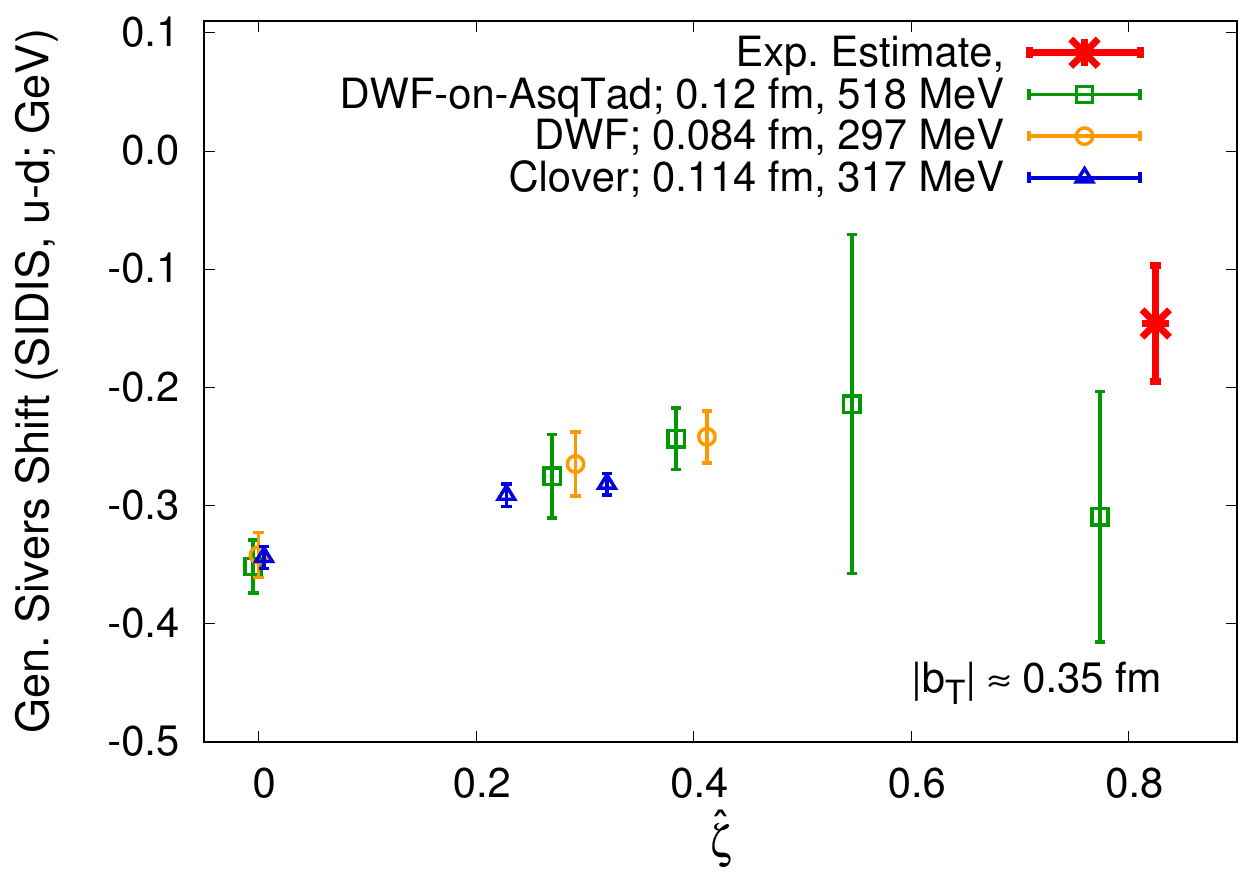}
\includegraphics[width=0.48\textwidth]{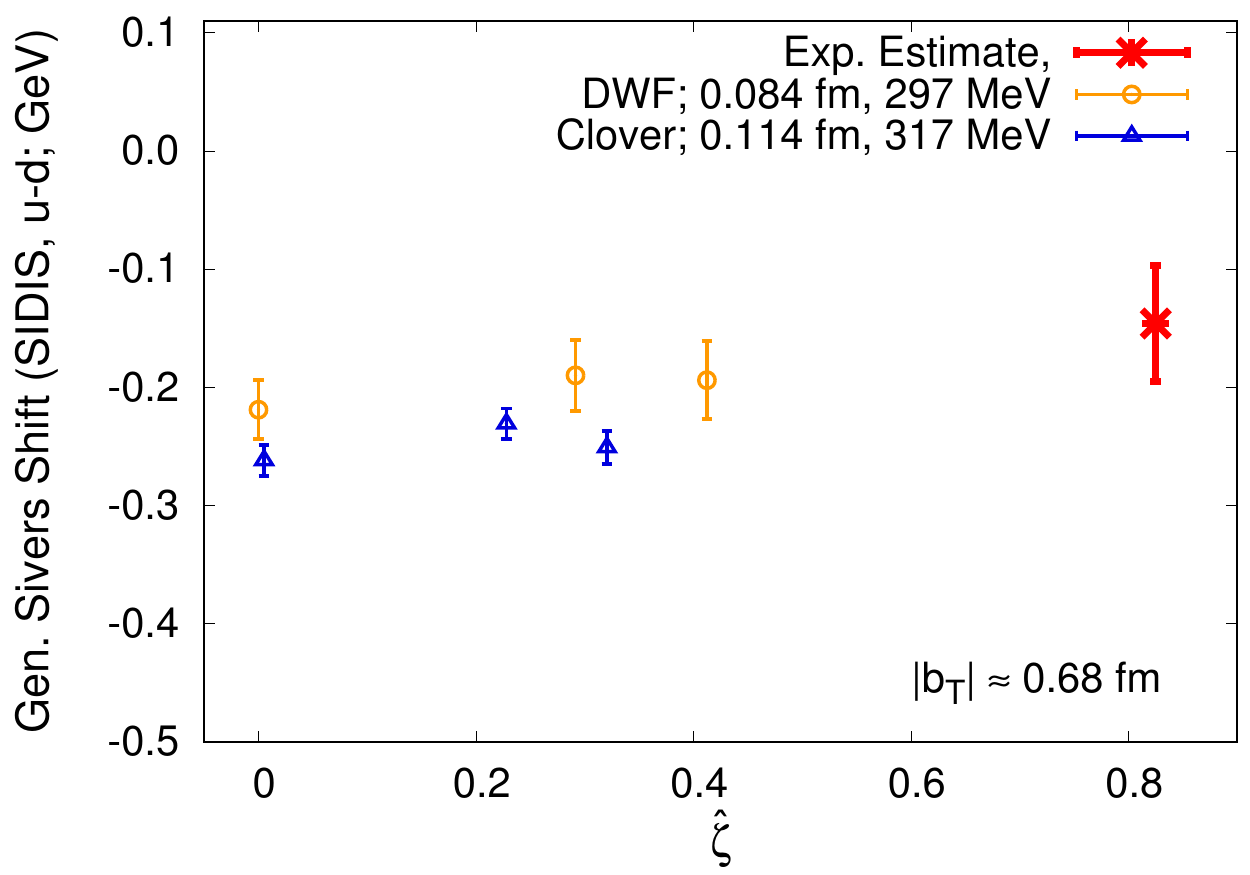}
\caption{Experimental extraction of the SIDIS generalized Sivers shift
  at $\zetahat=0.83$, together with Lattice QCD data in the SIDIS
  limit, $\eta \to \infty$, as a function of the Collins-Soper
  parameter $\hat{\zeta}$. Lattice data for $|\vprp{b}|\approx
  0.35$~fm are given in the left panel where we have included results
  from an earlier DWF-on-Asqtad study given in
  Ref.~\cite{Musch:2011er}. Results for 
  $|\vprp{b}| \approx 0.68$~fm are given in the right panel.  }
\label{fig:comp_exp}
\end{figure*}

In Fig.~\ref{fig:comp_exp}, we compare this result with lattice
estimates reproduced from Fig.~\ref{fig:sivers-evolX} for two values
of $|\vprp{b}| \approx 0.35$ and 0.68~fm. In the left panel, we also
include previous lattice results from a DWF-on-Asqtad study given in
Ref.~\cite{Musch:2011er}.  Note that the extraction of the
experimental estimate has been done at $\zetahat=0.83$, while precise
lattice results are obtained at $\zetahat \le 0.41$ (the earlier
lattice data points at $\zetahat > 0.41$ from Ref.~\cite{Musch:2011er}
have large uncertainties).  We observe the following:
First, the three lattice ensembles with different pion masses ($m_\pi
= 518 \MeV$ versus $m_\pi \approx 300 \MeV$) and different
discretization schemes at different values of the lattice spacing
give consistent results.
Second, as $|\vprp{b}|$ and/or $\hat{\zeta}$ are increased, the
lattice results tend toward the phenomenologically extracted value.
Third, the observed behavior is similar to that seen in the study
using pions in Ref.~\cite{Engelhardt:2015xja}. Thus, taking the trend
in our data between $0.2 < \hat{\zeta} < 0.41$ at face value, it is
reasonable to expect future lattice estimates at $\zetahat \approx
0.8$ to agree with the phenomenological value.

\section{Conclusion}
\label{sec:sum}

We present Lattice QCD results for the time-reversal odd generalized
Sivers and Boer-Mulders transverse momentum shifts applicable to SIDIS
and DY experiments; and for the T-even generalized transversity,
related to the tensor charge, and the generalized $g_{1T}$
worm-gear shift.  The lattice calculations were performed on two
different $n_f = 2+1$ flavor ensembles: a DWF ensemble with lattice
spacing $a=0.084$~fm and pion mass 297~MeV, and a clover ensemble with
$a=0.114$~fm and pion mass 317~MeV. The high statistics analysis of
the clover ensemble yields estimates with $O(10\%)$ uncertainty for
all four quantities over the range $|\vprp{b}| < 0.8$~fm and $\hat{\zeta}
\lesssim 0.3$. Estimates from the DWF ensemble have appreciably higher
statistical errors owing to the more limited statistics, but are
expected to have smaller systematic uncertainties.

Our results for TMD observables on two ensembles with comparable pion masses,
but with very different discretization of the Dirac action provide an
opportunity for an empirical test of the presence of finite lattice spacing
effects and the cancellation of renormalization factors in the ratios of
correlation functions considered. Estimates with DWF at $a=0.084$~fm are
expected to have small discretization errors. Apart from the notable
exception of the generalized $g_{1T}$
worm-gear shift, the consistency of DWF results with those using
clover fermions on coarser lattices with $a=0.114$~fm suggests
that lattice discretization effects are small.  

In continuum QCD, the nonlocal TMD operator is renormalized
multiplicatively with a renormalization factor composed of a product
of soft factors, operator specific, and quark wave function
renormalizations. This pattern is, a priori, not guaranteed to carry
over to the lattice formulation of the theory. Even though all the TMD
observables considered in the present work were calculated using
unrenormalized operators, the results for the ratios obtained using DW
and clover fermions are consistent except in some
specific circumstances. To the extent that they are consistent, this
can be taken as an indication that the renormalization factors largely 
cancel in the ratios considered.

The results for the TMD ratios obtained in the SIDIS and DY limits, i.e.,
using staple-shaped gauge connections, agree within uncertainties for
all four observables studied once the quark separation $|\vprp{b} |$
in the bilocal TMD operator exceeds about three lattice spacings. The
agreement furthermore persists into the regime of small $|\vprp{b} |$
for all but one of the TMD observables, namely, the generalized $g_{1T} $
worm-gear shift. Thus, within the statistical accuracy of the calculation, the
discretization effects and the cancellation of the renormalization
factors in our TMD observables in the SIDIS and DY limits appear under
control at finite physical separations $|\vprp{b} |$.

A surprising departure from the expectation that renormalization
factors generally become independent of the Dirac structure for
well-separated bilocal operators is observed for the T-even $g_{1T} $
worm-gear shift in the $\eta=0$ straight-link case. The discrepancy in
the $g_{1T} $ worm-gear shift at small $|\vprp{b} |$ for $\eta \to
\infty$ is seen to persist to all values of $|\vprp{b} |$ for $\eta =
0$. As discussed in Sec.~\ref{sec:straight}, we provide a plausible
explanation based on the recent 1-loop perturbative
calculation~\cite{Constantinou:2017sej} of a mixing, a lattice artifact in
the clover formulation, between axial and tensor operators for our
choice of the direction of the straight-link path vis-\`a-vis the
operator tensor index. Further studies that
include a complete analysis of the mixing, including a
non-perturbative calculation of the relevant renormalization factors,
are warranted to establish our observation.  Note that, whereas the
T-even functions with $\eta =0$ are not immediately relevant for TMD
applications, which call for staple-shaped gauge connections, such
operator mixing would need to be taken into account in the study of
PDFs
\cite{Ji:2013dva,Lin:2014zya,Alexandrou:2015rja,Ishikawa:2016znu,Chen:2016fxx,Chen:2016utp,Chen:2017mzz,Alexandrou:2016jqi,Alexandrou:2017huk},
which employ straight gauge connections.

Compiling the lattice TMD results obtained to date, as exhibited in
Fig.~\ref{fig:comp_exp} for the case of the Sivers shift, we observe that
three lattice ensembles with different pion masses and different
discretization effects give consistent results. In
an ideal case, in which estimates are obtained with arbitrarily small
errors, such a consistency could be taken as evidence that the
dependence on the light quark masses and the discretization
corrections are both small.  Furthermore, as discussed in Sec.~\ref{sec:setup}, 
the renormalization factors cancel in the ratio defining the Sivers
shift. We therefore regard it as reasonable to compare lattice results
obtained to date for the Sivers shift at pion masses down to $m_{\pi }
\approx 300$~MeV to a phenomenological estimate extracted from
experimental data. Indications of consistency with the experimental
result at $\hat{\zeta} \gtrsim 0.8$, cf.~Fig.~\ref{fig:comp_exp},
suggest that, within our uncertainties, lattice artifacts are already
reasonably small at the values of the lattice parameters
employed. Future higher precision calculations on ensembles with
lighter quark masses are, therefore, well-motivated.

\section*{Acknowledgments}
We thank Martha Constantinou, Zhongbo Kang and Stefan Meinel for
fruitful discussions. The RBC/UKQCD collaboration is gratefully
acknowledged for providing the DWF ensemble analyzed in this work, as
are R.~Edwards, B.~Jo\'{o}, and K.~Orginos for providing the clover
ensemble, which was generated using resources provided by XSEDE
(supported by National Science Foundation Grant
No.~ACI-1053575). Computations were performed using resources provided
by the U.S.~DOE Office of Science through the National Energy Research
Scientific Computing Center (NERSC), a DOE Office of Science User
Facility, under Contract No. DE-AC02-05CH11231, as well as through
facilities of the USQCD Collaboration, employing the Chroma software
suite \cite{Edwards:2004sx}.  The work of T.~Bhattacharya, R.~Gupta
and B.~Yoon is supported by the U.S. Department of Energy, Office of
Science, Office of High Energy Physics under contract number
DE-KA-1401020 and the LANL LDRD program.  M.~Engelhardt, J.~Negele and
A.~Pochinsky are supported by the U.S. Department of Energy, Office of
Science, Office of Nuclear Physics through grants numbered DE-FG02-
96ER40965, DE-SC-0011090 and DE-FC02-06ER41444 respectively.
S.~Syritsyn was supported by the U.S. Department of Energy, Office of
Science, Office of Nuclear Physics under contract DE-AC05-06OR23177
and by the RIKEN BNL Research Center under its joint tenure track
fellowship with Stony Brook University. A.~Sch\"afer is supported by DFG (SFB-TRR 55).

\appendix
%

%
\bibliographystyle{apsrev} 
\bibliography{ref} 

\end{document}